\documentclass[english,13pt]{article}
\usepackage[T1]{fontenc}
\usepackage[latin1]{inputenc}
\usepackage{geometry}
\usepackage{float}
\usepackage[section]{placeins}
\usepackage{mathrsfs}
\usepackage{amsmath}
\usepackage{amssymb}
\usepackage{graphicx}
\usepackage{booktabs}
\usepackage{multirow}
\usepackage{hyperref}
\hypersetup{
    colorlinks=true,
    linkcolor=blue,
    filecolor=magenta,
    urlcolor=cyan,
    citecolor = red,
}

\makeatletter

\usepackage{algorithm,algpseudocode}

\usepackage{amsthm}

\usepackage{mathrsfs}

\usepackage{amsfonts}

\usepackage{epsfig}

\usepackage{bm}

\usepackage{mathrsfs}

\usepackage{enumerate}

\numberwithin{equation}{section}

\@ifundefined{definecolor}{\@ifundefined{definecolor}
 {\@ifundefined{definecolor}
 {\usepackage{color}}{}
}{}
}{}

\usepackage{subfig}\usepackage[all]{xy}

\newtheorem{prop}{Proposition}[section]

\newcounter{hypA}
\newenvironment{hypA}{\refstepcounter{hypA}\begin{itemize}
  \item[({\bf A\arabic{hypA}})]}{\end{itemize}}
\newcounter{hypB}

\newcounter{hypD}

\newcounter{hypW}

\usepackage{babel}\date{}

\usepackage{babel}

\makeatother

\usepackage{babel}

\begin{document}

\begin{center}

{\Large \textbf{Sequential Markov Chain Monte Carlo for
Filtering of State-Space Models with Low or Degenerate Observation Noise}}

\vspace{0.5cm}

BY  ABYLAY ZHUMEKENOV$^{1}$,  ALEXANDROS BESKOS$^{2}$,  DAN CRISAN$^{3}$,
AJAY JASRA$^{1}$ \& NIKOLAS KANTAS$^{3}$

{\footnotesize $^{1}$School of Data Science,  The Chinese University of Hong Kong,  Shenzhen,  Shenzhen, CN.}\\
{\footnotesize $^{2}$Department of Statistical Science, University College London, London, WC1E 6BT, UK}\\
{\footnotesize $^{3}$Department of Mathematics, Imperial College London, London, SW7 2AZ, UK}\\
{\footnotesize E-Mail:\,} \texttt{\emph{\footnotesize  abylayzhumekenov@cuhk.edu.cn;
a.beskos@ucl.ac.uk;d.crisan@ic.ac.uk;
ajayjasra@cuhk.edu.cn; n.kantas@ic.ac.uk
}}

\end{center}

\begin{abstract}
We consider the discrete-time filtering problem in scenarios where the observation noise is degenerate or low. More precisely, one is given access to a discrete time observation sequence which at any time $k$ depends only on the state of an unobserved Markov chain. We specifically assume that the functional relationship between observations and hidden Markov chain has either degenerate or low noise. In this article, under suitable assumptions, we derive the filtering density and its recursions for this class of problems on a specific sequence of manifolds defined through the observation function. We then design sequential Markov chain Monte Carlo methods to approximate the filter serially in time. For a certain linear observation model, we show that using sequential Markov chain Monte Carlo for low noise will converge as the noise disappears to that of using sequential Markov chain Monte Carlo for degenerate noise. We illustrate the performance of our methodology on several challenging stochastic models arising in statistics and applied mathematics.
\\
\newline
\noindent \textbf{Key words}: Observation Noise, Filtering,  Manifold Markov Chain Monte Carlo.
\end{abstract}

\section{Introduction}

We consider the following filtering problem (e.g.~\cite{bain2009fundamentals}).  We have access to discrete-time observations $Y_1,Y_2,\dots$,
$Y_k\in\mathbb{R}^{d_y}$,  $k\in\mathbb{N}$,
 that are associated to an unobserved discrete-time Markov chain $X_0,X_1,\dots$,  $X_k\in\mathbb{R}^{d_x}$,  $k\in\mathbb{N}$.  It is supposed that at any
time $k\in\mathbb{N}$ we can write either
\begin{align*}
Y_k & =h_k(X_k), \quad \text{degenerate noise,}\\
Y_k & =\widetilde{h}_k(X_k,\epsilon_k),\quad \text{low noise,}
\end{align*}
where $h_k:\mathbb{R}^{d_x}\rightarrow\mathbb{R}^{d_y}$ is not injective and $\epsilon_1,\epsilon_2,\dots$
is an independent and identically distributed (i.i.d.) sequence of $\mathbb{R}^{d_y}$ random variables with `small variance',  $\widetilde{h}_k:\mathbb{R}^{d_x}\times\mathbb{R}^{d_y}\rightarrow\mathbb{R}^{d_y}$.
To substantiate the phrase small variance,  one could imagine that $\epsilon_k$ has a $d_y-$dimensional Gaussian distribution of mean zero and covariance $\Delta\mathtt{I}_{d_y}$ with $\Delta>0$ but close to zero and $\mathtt{I}_{d_y}$ the $d_y\times d_y$ identity matrix.
The objective of either the degenerate or low (or small) noise case,
assuming it is well-defined,  is to estimate expectations with respect to (w.r.t.) the conditional distribution of $X_k$ given all the observations up-to time $k$,  $Y_{1:k}=(y_1,\dots,y_k)^{\top}$, recursively in time.  These problems
are important in many applications in filtering problems where it is expected that the signal dynamics (unobserved Markov chain) is strongly driving the observed data; see for instance \cite{barczyk2015invariant, hua2013implementation, srivastava2004bayesian} for several applications.

Understanding filtering in the low noise regime has attracted long interest in the past from researchers focusing on continuous time dynamics and devising robust (extended) Kalman-Bucy type approximations of the filter \cite{brigo1996new, katzur1984asymptotic1, de1994approximate, picard1986nonlinear, picard1991efficiency} or establishing large deviations results for low $\Delta$  \cite{hijab1984asymptotic, pardoux2004quenched}, see also \cite{kutoyants2025extended} for a recent review. In general, we need to stress that the filtering distributions are typically analytically intractable except in fairly simple models,  such as ones involving linear and Gaussian dynamics.  Using analytical approximations such as the extended Kalman filter or variants are biased and most often do not perform well for many models. As a result,  there is a vast literature in designing numerical algorithms based on Monte Carlo for approximating the filter; we refer the reader to \cite{bain2009fundamentals, moral2004feynman} and the references therein for a complete treatment.  In this article we focus upon developing Monte Carlo based algorithms for approximating the filter for the degenerate or low noise case.  Perhaps the most popular simulation method for low-dimensional filtering problems ($d_x\in\{1,2,\dots,10\}$) is the particle filter (PF)
(see e.g.~\cite{bain2009fundamentals, moral2004feynman}) which entails sampling $N\in\mathbb{N}$ particles in parallel,  with the particles undergoing sampling,  weighting and resampling operations.  There are several related methods in high-dimensions such as
\cite{beskos2017stable, ruzayqat2022lagged} and the approach that we are to focus upon is sequential Markov chain Monte Carlo (SMCMC) \cite{berzuini1997dynamic, centanni2006monte, martin2013inference} which has been shown to be one of the most competitive methods (at least empirically) when trying to approximate the filter in high-dimensions \cite{ruzayqat2024sequential}.

The main issue that we deal with in this article is the appropriate formulation of the filter in a manner that takes into consideration the observation regime of interest and the corresponding development of effective computational approximations.  In the degenerate noise case,  we follow works such as \cite{diaconis2013sampling, graham2017asymptotically, graham2022manifold} in defining the filter on a manifold $\mathsf{M}_k=\{x\in\mathbb{R}^{d_x}:y_k-h_k(x)=0\}$.  In general,  this manifold has zero Lebesgue measure and one must seek to derive a filter w.r.t.~a more appropriate dominating measure.
Under appropriate mathematical conditions we are able to derive the filtering density and the traditional filtering recursions of this density w.r.t.~the Riemannian measure,  i.e.~we focus on the case where we can perform filtering on Riemannian manifolds. Previously, \cite{joannides1995nonlinear} also considered defining the filter along these lines, but focused only on continuous time dynamics for $X$ and did not provide filtering recursions useful for discrete time models or appropriate numerical methods, which is the focus of this paper.
We further show that the low-noise case can be cast in a similar framework and this can have substantial benefits in designing efficient numerical methods that are robust to very low observation noise.
Having formulated the filter and its recursions we then seek to design computational methods for its approximation.  Generic particle filtering or SMCMC methodologies referenced in the previous paragraph are not designed for filtering with degenerate noise.
In particular, using a particle filter in the setting of interest in this work gives rise to some challenges:
\begin{enumerate}
\item{In the context of degenerate noise one has to design importance sampling proposal dynamics that can move from one manifold to another.}
\item{In the small noise case,  one expects that standard implementations of particle filters, oblivious to the  prescribed model structure, will be rather inefficient},  with a likely collapse of the importance weights to zero.
\end{enumerate}
Whilst 1.~is certainly an interesting open research question,  it requires a substantial effort and then, at best,  one can address low dimensional filtering problems.  In 2.~this circumvents the use of the PF in small noise filtering problems.

The difficulty in the degenerate case is structural rather than a matter of proposal design. For any proposal $q(x_k|x_{k-1}^i,y_k)$ that is absolutely continuous with respect to the Lebesgue measure, the probability that a proposed particle lands exactly on the constraint manifold $\mathsf{M}_k$ is zero. Consequently, all importance weights vanish and the filter collapses, regardless of whether one uses a standard bootstrap filter, an auxiliary particle filter \cite{pitt1999filtering}, an implicit particle filter \cite{chorin2010implicit}, or a twisted particle filter \cite{whiteley2014twisted}. This limitation has motivated a number of alternative approaches that incorporate the observational constraint directly into the dynamics, such as guided SDEs based on the Doob $h$-transform \cite{schauer2017guided,chopin2023computational,pieper2025guided}, barrier-function modifications of the drift \cite{erdogan2025class}, and ensemble Kalman filters \cite{evensen2003ensemble}. In the special case of linear observation operators, a particle filter that operates directly on the manifold was recently proposed in \cite{zhumekenov2026particle}. Ideally, we want a method that requires no bespoke proposal design or approximate dynamics, and targets the Riemannian measure on the manifold directly. We do not claim that SMCMC is universally preferable to particle filtering, but rather that it is well suited to the degenerate and low-noise observation regimes considered here.

More recently, score-based generative models have been employed to directly approximate filtering distributions, circumventing the need for explicit proposal design or manifold parameterisations \cite{bao2025nonlinear, millard2026particle}. These methods represent an emerging alternative to the MCMC-based approach, but require pre-training and running reverse diffusion, both of which are slow and expensive.

In the context of using MCMC for sampling from probability measures defined on a manifold,  there is already a substantial number of methods; a non-exhaustive list includes \cite{bharath2025sampling, byrne2013geodesic, graham2017asymptotically, graham2022manifold, zappa2018monte}.  Given the success of the application of SMCMC in filtering in \cite{ruzayqat2024sequential} it therefore seems natural to then design SMCMC for our class of filtering problems.

The contribution of the present work lies in the principled combination of existing components into a general-purpose filtering framework that requires no custom proposal design, applies to any smooth constraint with full-rank Jacobian, and maintains constant per-step cost.

The main contributions of this article are as follows:
\begin{itemize}
\item{Establishing a sequence of filtering densities w.r.t.~the Riemannian measure,  along with filtering recursions for the degenerate noise case and then extending this framework to encompass the low noise case.}
\item{Developing SMCMC methods that combine the SMCMC framework with a general-purpose manifold MCMC kernel, requiring no custom proposal design per constraint type.}
\item{In a particular scenario,  proving that SMCMC in the small noise case will converge as the noise goes to zero,
(marginally) to SMCMC in the degenerate noise problem. }
\item{Implementing our methodology on several interesting models.}
\end{itemize}

Given that we can design efficient methods for filtering in the degenerate noise case,  by casting the low noise problem into a similar framework,  with little extra effort we can carry out filtering for an arbitrarily low noise. In fact this approach will be robust to low levels of $\Delta$ and the convergence result mentioned in the third bullet means that this is an appropriate strategy when sampling for the limiting degenerate case performs well. We stress that standard algorithms do not satisfy such a robustness criterion w.r.t. diminishing noise.
This extends the appeal of the mathematical and computational approach in this article. Finally it is worth mentioning that there are problems where $X_k$ is naturally defined on a manifold and this requires one to establish appropriate filtering equations, e.g. \cite{crisan2009nonlinear, zhang2017feedback} consider this in continuous-time. Here our setting is different and we establish filtering recursions on a sequence of manifolds $(\mathsf{M}_k)$ due to the constraint posed by the observations.

This article is structured as follows.  In Section \ref{sec:filter} we define the filtering problem of interest, which features
our first mathematical result on defining the sequence of filtering probability measures.
In Section~\ref{sec:method} we present our sequential MCMC method.
This section also gives the second mathematical result on proving that SMCMC in the small noise case will converge
to SMCMC in the degenerate noise problem.
In Section \ref{sec:numerics} we detail
our numerical implementations.   The proofs of our mathematical results can be found in Appendix \ref{app:proofs}.

\section{Filtering Problem}\label{sec:filter}

\subsection{Model}

We denote the measurable space of the $d_x-$dimensional real line with associated Borel $\sigma-$field
$(\mathbb{R}^{d_x},\mathcal{B}(\mathbb{R}^{d_x}))$ and set $\lambda_{d_x}$ as the $d_x-$dimensional Lebesgue measure.
We are given a discrete-time Markov Chain $\{X_k\}_{k\geq 0}$,  $X_k\in\mathbb{R}^{d_x}$,   with $X_0=x_0\in\mathbb{R}^{d_x}$ given and positive Lebesgue transition density $f_k:\mathbb{R}^{d_x}\times\mathbb{R}^{d_x}\rightarrow\mathbb{R}^+$, $k\in\mathbb{N}$.  That is,  for any $(k,x_{k-1})\in\mathbb{N}\times\mathbb{R}^{d_x}$:
$$
\int_{\mathbb{R}^{d_x}}f_k(x_{k-1},x_k)\lambda_{d_x}(dx_k) = 1
$$
and any $(k,A)\in\mathbb{N}\times\mathcal{B}(\mathbb{R}^{d_x})$
$$
\int_A f_k(x_{k-1},x_k)\lambda_{d_x}(dx_k)
$$
is a measurable function.
For any fixed $n\in\mathbb{N}$, the joint density, of the Markov chain w.r.t.~$\bigotimes_{k=1}^n \lambda_d(dx_k)$ is
$$
p_n(x_{1:n}) = \prod_{k=1}^n f_k(x_{k-1},x_k).
$$

We consider a sequence of observations $\{Y_k\}_{k\geq 1}$,  $Y_k\in\mathbb{R}^{d_{y}}$ which are assumed such that for $h_k:\mathbb{R}^{d_x}\rightarrow\mathbb{R}^{d_y}$, $k\in\mathbb{N}$, we have
$$
Y_k = h_k(X_k)
$$
where $h_k$ is not injective.
The objective,  assuming it is well-defined,  is to approximate expectations
of functions of $X_k$ given all the observations $y_{1:k}$ at each time $k\in\mathbb{N}$ (i.e.~filtering).   We refer to this problem as `filtering with degenerate noise'.
Throughout we will only be able to deal with models for which $d_x>d_y$ as we will define our filter on
a $d_x-d_y$ submanifold of $\mathbb{R}^{d_x}$,  with $\mathbb{R}^{d_x}$ itself being the ambient space.

We will also consider the case that
\begin{equation}\label{eq:low_noise}
Y_k=\widetilde{h}_k(X_k,\epsilon_k)
\end{equation}
where
$\widetilde{h}_k:\mathbb{R}^{d_x}\times\mathbb{R}^{d_y}\rightarrow\mathbb{R}^{d_y}$ where $\{\epsilon_k\}_{k\geq 0}$
is an independent and identically distributed (i.i.d.) sequence of noises which have zero mean and small variance; filtering in this context will be termed `filtering with low noise'.  As we will see,  this latter problem can be dealt with via an approach similar to the one developed for the degenerate case.

\subsection{Filtering with Degenerate Noise}

For each $k\in\mathbb{N}$ we write
$$
c_k(x_k) = y_k-h_k(x_k).
$$
For any fixed $n\in\mathbb{N}$ write $\mathbf{c}_n(x_{1:n}):=(c_1(x_1),\dots,c_n(x_n))^{\top}$.
Introduce the manifold $\mathsf{M}_k=\{x\in\mathbb{R}^{d_x}:c_k(x)=0\}$ and set $\pmb{\mathsf{M}}_n
=\{x_{1:n}\in\mathbb{R}^{nd_x}:\mathbf{c}_k(x_{1:n})={0}\}=\mathsf{M}_1\times\cdots\times\mathsf{M}_n$.
Let $\{M_k\}_{k\geq 1}$ be any sequence of positive definite $d_x\times d_x$ matrices and we suppose that for each $k\in\mathbb{N}$,  $\mathbb{R}^{d_x}$ is equipped with a metric tensor with a fixed positive definite matrix representation $M_k$.  For $n\in\mathbb{N}$ set $\mathbf{M}_n=\text{diag}(M_1,\dots,M_n)$ be a $nd_x\times n d_x$ block diagonal matrix.  We suppose that for each $n\in\mathbb{N}$, $\mathbb{R}^{nd_x}$ is equipped with a metric tensor with a fixed positive definite matrix representation $\mathbf{M}_n$.
We will make the following assumption.

\begin{hypA}\label{ass:1}
For each $k\in\mathbb{N}$,  $c_k$ is continuously differentiable and that the Jacobian $\partial  c_k$ has full row-rank.
\end{hypA}

For $k\in\mathbb{N}$,  we denote by $\sigma_{\mathsf{M}_k}^{M_k}(dx_k)$ the Riemannian measure on the manifold $\mathsf{M}_k$;  see for instance \cite[Lemma C.2.]{graham2022manifold}. In effect, $\sigma_{\mathsf{M}_k}^{M_k}(dx_k)$ is the natural extension of a Lebesgue measure on a manifold, in the sense that, e.g., it can be used to measure surfaces on the manifold.
For simplicity we write for $k\in\mathbb{N}$
\begin{equation}
\label{eq:g-function}
g_k(x_k) = \mathrm{det}\left(\partial c_k(x_k)M_k^{-1}\partial c_k(x_k)^{\top}\right)^{-1/2}.
\end{equation}
We will denote by $\pi_k$ the filtering density, w.r.t.~$\sigma_{\mathsf{M}_k}^{M_k}$ at any time $k\in\mathbb{N}$ .   We write the density as $\pi_k(x_k)$,  that is,  omitting making reference to the data in the notation.
Our objective is estimate expectations w.r.t.~the filter,  that is for $\varphi:\mathbb{R}^{d_x}\rightarrow\mathbb{R}$ to estimate
$$
\pi_k(\varphi) =\mathbb{E}[\varphi(X_k)|Y_{0:k}]= \int_{\mathsf{M}_k}\varphi(x_k)\pi_k(x_k)\sigma_{\mathsf{M}_k}^{M_k}(dx_k)
$$
when $\varphi$ is integrable w.r.t.~$\pi_k(x_k)\sigma_{\mathsf{M}_k}^{M_k}(dx_k)$.
In the following result we give a representation of the density $\pi_k$. The proof can be found in Appendix \ref{app:app1}.

\begin{prop}\label{prop:prop1}
Assume (A\ref{ass:1}).  Then we have that
$$
\pi_1(x_1) = \frac{g_1(x_1) f_1(x_0,x_1)}
{
\int_{\mathsf{M_1}}
g_1(x_1)
f_1(x_0,x_1)
\sigma_{\mathsf{M}_1}^{M_1}(dx_1)
}
$$
and for any $k\geq 2$ we have that
\begin{equation}\label{eq:pred_up}
\pi_k(x_k) =
\frac{
g_k(x_k)
\int_{\mathsf{M_{k-1}}}
f_k(x_{k-1},x_k)
\pi_{k-1}(x_{k-1})
\sigma_{\mathsf{M}_{k-1}}^{M_{k-1}}(dx_{k-1})
}
{
\int_{\mathsf{M_k}}
g_k(x_k)
\int_{\mathsf{M_{k-1}}}
f_k(x_{k-1},x_k)
\pi_{k-1}(x_{k-1})
\sigma_{\mathsf{M}_{k-1}}^{M_{k-1}}(dx_{k-1})
\sigma_{\mathsf{M}_k}^{M_k}(dx_k)
}.
\end{equation}
Functions $g_1, g_2,\ldots$ are as defined in (\ref{eq:g-function}).
\end{prop}

The significance of Proposition \ref{prop:prop1} is to demonstrate that under fairly general conditions one has a prediction-updating representation of the filtering density w.r.t.~the Riemannian measure,  in the case of degenerate noise.  This formula is not surprising,  given the previous work in e.g.~\cite{diaconis2013sampling, graham2022manifold},  but the formal statement is important in the computational methodology that we are to present.  In essence we seek to construct a computational approach which can approximate this prediction-updating formula \eqref{eq:pred_up}.

We note that the formulation of \cite{graham2022manifold} considers the setting where $X_k=F_k(X_{k-1},\nu_k)$ for $\{\nu_k\}_{k\geq 0}$ a sequence of i.i.d.~noises with positive Lebesgue density and $F_k:\mathbb{R}^{2d_x}\rightarrow\mathbb{R}^{d_x}$.  Under appropriate assumptions this allows one to define a type of filter on the \emph{path-space} of noises $\nu_{1:k}$ at any time $k$.  The details can be found in \cite{graham2022manifold},  but for the purpose of this discussion suppose given $x_0$ one has that $X_1=F_1(x_0,\nu_1)$ and then for any subsequent $k\geq 2$,  $X_k= F_k(F_{k-1}\circ\cdots\circ F_1(x_0,\nu_1),\nu_k)$, which then induces a constraint $\mathbf{c}_k(\nu_{1:k})=0$.
The motivation from \cite{graham2022manifold} is that at each $k$ this so-called collapsed representation (as per~\cite{murray2013disturbance}) of the smoother/filter can facilitate MCMC algorithms targeting $\nu_{0:k}$, which can mix quickly, rather than working directly with the $X_k$'s.  In the context of filtering,  this approach can have two deficiencies:
\begin{enumerate}
\item{The state-space of the filter grows with time.}
\item{The number of constraints grows with time.}
\end{enumerate}
Point 1.~is well understood in the literature (e.g.~\cite{kantas2014sequential}) and often leads to computational algorithms whose cost scales quadratically with the time parameter.  In online problems (i.e.~filtering) one often wants the computational cost to be fixed each time an observation is acquired so that long data sequences can be processed.  On point 2.~the issue is that when considering simulation methods to update all the $\nu_{1:k}$ one has to satisfy all of the constraints $\mathbf{c}_k(\nu_{1:k})$ simultaneously,  which is possible,  e.g.~with MCMC \cite{byrne2013geodesic, graham2017asymptotically, graham2022manifold, zappa2018monte},   but, unquestionably would add extra computational complexity to a simulation method.
As a result,  we have decided to work directly with the $X_k$'s as in Proposition \ref{prop:prop1}.

\subsection{Filtering with Low Noise}

We now consider the case \eqref{eq:low_noise}.
Filtering with low noise of course, under appropriate assumptions, can be dealt with via a density w.r.t.~$d_x-$dimensional Lebesgue
measure.  However,  in a manner similar to the approach in \cite{graham2022manifold} one can also consider an equivalent representation w.r.t.~the Riemannian measure.  We will explain the possible benefits from opting for such an approach when the noise has small variance below.

We give a formal presentation as above; the details are repeated for completeness.  For each $k\in\mathbb{N}$ we write
$$
\widetilde{c}_k(x_k,\epsilon_k) = y_k-\widetilde{h}_k(x_k,\epsilon_k).
$$
For any fixed $n\in\mathbb{N}$ write $\widetilde{\mathbf{c}}_n(x_{1:n},\epsilon_{1:n}):=(\widetilde{c}_1(x_1,\epsilon_1),\dots,\widetilde{c}_n(x_n,\epsilon_n))^{\top}$.
Set $\widetilde{\mathsf{M}}_k=\{(x,\epsilon)\in\mathbb{R}^{d_x+d_y}:\widetilde{c}_k(x,\epsilon)=0\}$ and
$\pmb{\widetilde{\mathsf{M}}}_n
=\{(x_{1:n},\epsilon_{1:n})\in\mathbb{R}^{n(d_x+d_y)}:\widetilde{\mathbf{c}}_n(x_{1:n},\epsilon_{1:n}):=0\}=\widetilde{\mathsf{M}}_1\times\cdots\times\widetilde{\mathsf{M}}_n$.
Let $\{\widetilde{M}_k\}_{k\geq 1}$ be any sequence of positive definite $(d_x+d_y)\times (d_x+d_y)$ matrices and we suppose that for each $k\in\mathbb{N}$,  $\mathbb{R}^{2d_x}$ is equipped with a metric tensor with a fixed positive definite matrix representation $\widetilde{M}_k$.  For $n\in\mathbb{N}$ set $\widetilde{\mathbf{M}}_n=\text{diag}(\widetilde{M}_1,\dots,\widetilde{M}_n)$ be the $n(d_x+d_y)\times n(d_x+d_y)$ block diagonal matrix.  We suppose that for each $n\in\mathbb{N}$, $\mathbb{R}^{n(d_x+d_y)}$ is equipped with a metric tensor with a fixed positive definite matrix representation $\widetilde{\mathbf{M}}_n$.   We assume that:
$p_{\texttt{N}}$ is the positive Lebesgue density of the distribution of $\epsilon_1$;  for each $k\in\mathbb{N}$,  $ \widetilde{c}_k$
is continuously differentiable
and that the Jacobian $\partial \widetilde{{c}}_k$ has full row-rank. Then one can prove in a similar manner to Proposition \ref{prop:prop1} that the filtering density at time 1 w.r.t.~$\sigma^{\widetilde{M}_1}_{\widetilde{\mathsf{M}}_1}$ is
$$
\widetilde{\pi}_1(x_1,\epsilon_1) = \frac{\widetilde{g}_1(x_1,\epsilon_1)
f_1(x_0,x_1) p_{\texttt{N}}(\epsilon_1)
}
{
\int_{\widetilde{\mathsf{M}}_1}
\widetilde{g}_1(x_1,\epsilon_1)
f_1(x_0,x_1) p_{\texttt{N}}(\epsilon_1)
\sigma_{\widetilde{\mathsf{M}}_1}^{\widetilde{M}_1}\left(d(x_1,\epsilon_1)\right)
}
$$
and for any $k\geq 2$ we have that
$$
\widetilde{\pi}_k(x_k,\epsilon_k) =
\frac{
\widetilde{g}_k(x_k,\epsilon_k)
p_{\texttt{N}}(\epsilon_k)
\int_{\widetilde{\mathsf{M}}_{k-1}}
f_k(x_{k-1},x_k)
\widetilde{\pi}_{k-1}(x_{k-1},\epsilon_{k-1})
\sigma_{\widetilde{\mathsf{M}}_{k-1}}^{\widetilde{M}_{k-1}}\left(d(x_{k-1},\epsilon_{k-1})\right)
}
{
\int_{\widetilde{\mathsf{M}}_k}
\widetilde{g}_k(x_k,\epsilon_k)
p_{\texttt{N}}(\epsilon_k)
\int_{\widetilde{\mathsf{M}}_{k-1}}
f_k(x_{k-1},x_k)
\widetilde{\pi}_{k-1}(x_{k-1},\epsilon_{k-1})
\sigma_{\widetilde{\mathsf{M}}_{k-1}}^{\widetilde{M}_{k-1}}\left(d(x_{k-1},\epsilon_{k-1})\right)
\sigma_{\widetilde{\mathsf{M}}_k}^{\widetilde{M}_k}\left(d(x_k,\epsilon_k)\right)
}
$$
where
$$
\widetilde{g}_k(x_k,\epsilon_k) =
\mathrm{det}\left(\partial \widetilde{c}_k(x_k,\epsilon_k)\widetilde{M}_k^{-1}\partial \widetilde{c}_k(x_k,\epsilon_k)^{\top}\right)^{-1/2}.
$$

The reason for this presentation is as follows.  If one has a computational method that can efficiently approximate the
filter in the degenerate noise case (i.e.~$\pi_k$),  then there is little complication to efficiently approximate the filter in the low noise case (i.e.~$\widetilde{\pi}_k$) and moreover there may be some type of convergence as the noise in the latter case disappears. More concretely,  suppose that $\epsilon_k\stackrel{\text{i.i.d.}}{\sim}\mathcal{N}_{d_y}(0,\Delta\mathtt{I}_{d_y})$,  where $\Delta>0$ and $\mathcal{N}_{d_y}(0,\Delta\mathtt{I}_{d_y})$ is the $d_y-$dimensional Gaussian distribution of mean 0 and covariance $\Delta$ multiplied by the identity matrix $\mathtt{I}_{d_y}$.  Then we might expect that if there is a computational method approximating $\{\widetilde{\pi}_k\}_{k\geq 0}$ then as $\Delta\downarrow 0$ there is a mathematical convergence of this method (the notion of convergence is left deliberately vague, but is explored in Section \ref{sec:conv_res}) to some type of equivalent in the degenerate noise case.  We note that popular sampling methods could fail to perform well in the low noise case.  For instance, one might expect quite sophisticated MCMC methods,  such as \cite{andrieu2010particle},  to have very poor mixing.  Finally, the presentation for the low noise case includes also intermediate cases where $\widetilde{h}_k(X_k,\epsilon_k)$ results to some coordinates of $Y_k$ being deterministic and others containing some level of noise.

\section{Methodology}\label{sec:method}

In the following section we outline our methodology.  All of the approaches are detailed in the degenerate noise case, but there is little modification needed for the low noise case.  At the end, Section \ref{sec:conv_res},  we give our result
on the convergence of SMCMC in the low noise case,  to that in the degenerate case.

\subsection{Basic Method}

Our approach is to use sequential MCMC (SMCMC) which proceeds as follows.  At time 1,  one needs access to any ergodic Markov kernel $K_1:\mathbb{R}^{d_x}\times\mathcal{B}(\mathbb{R}^{d_x})\rightarrow[0,1]$ that admits $\pi_1(x_1)\sigma_{\mathsf{M}_1}^{M_1}(dx_1)$ as its invariant measure.   Several candidates are available such as in \cite{byrne2013geodesic, graham2017asymptotically, graham2022manifold, zappa2018monte}
and for completeness we detail the approach of \cite{zappa2018monte} in Section \ref{sec:ex_mcmc}.

The MCMC kernel will produce $N$ samples which we denote $(x_1^1,\dots,x_1^N)\in\mathsf{M}_{1}^{N}$.
Now,  the filter at time 2 as given in Proposition \ref{prop:prop1} is
$$
\pi_2(x_2) =
\frac{
g_2(x_2)
\int_{\mathsf{M_{1}}}
f_2(x_{1},x_2)
\pi_{1}(x_{1})
\sigma_{\mathsf{M}_{1}}^{M_{1}}(dx_{1})
}
{
\int_{\mathsf{M_2}}
g_2(x_2)
\int_{\mathsf{M_{1}}}
f_2(x_{1},x_2)
\pi_{1}(x_{1})
\sigma_{\mathsf{M}_{1}}^{M_{1}}(dx_{1})
\sigma_{\mathsf{M}_2}^{M_2}(dx_2)
}
$$
which for any given $x_2$ is not available to compute point-wise up-to a normalizing constant; this is a pre-requisite of using MCMC, which is our intention.  The idea in SMCMC is to use the samples at the previous observation time, to approximate the density of the target at the current observation time.  In this scenario,  one would have the approximated density
$$
\pi_2^N(x_2) \propto g_2(x_2)\frac{1}{N}\sum_{i=1}^N
f_2(x_{1}^i,x_2)
$$
w.r.t.~$\sigma_{\mathsf{M}_2}^{M_2}$.  At this stage one may be tempted to sample from $\pi_2^N$,  but this can be computationally prohibitive as the cost of computing $\pi_2^N$ is $\mathcal{O}(N)$.  Since one typically must compute $\pi_2^N$ to simulate from an MCMC kernel of invariant measure
$\pi_2^N(x_2)\sigma_{\mathsf{M}_2}^{M_2}(dx_2)$,  the cost to produce $N$ samples is $\mathcal{O}(N^2)$ which can be too much for the approach to be practically useful.

We consider the following simple alternative.  Let $s\in\{1,\dots,N\}$ be given and typically we want $s$ to be much less than $N$.  Define $\mathsf{I}_s=\{i_{1:s}\in\{1,\dots,N\}^s:i_1\neq i_2\neq\cdots\neq i_s\}$
(write the power set of $\mathsf{I}_s$ as $\mathcal{I}_s$),  we then define the probability density
$$
\widehat{\pi}_2^N(x_2,i_{1:s}) \propto g_2(x_2)\sum_{j\in\mathsf{I}_s}
f_2(x_{1}^j,x_2)
$$
w.r.t.~$\sigma_{\mathsf{M}_2}^{M_2}\otimes\mu_{\mathsf{I}_s}$,  where $\mu_{\mathsf{I}_s}$ is the counting measure on $\mathsf{I}_s$.  We note that it is straightforward to verify that:
$$
\pi_2^N(x_2) = \sum_{i_{1:s}\in\mathsf{I}_s}
\widehat{\pi}_2^N(x_2,i_{1:s})
$$
so that one can sample directly from $\widehat{\pi}_2^N$ and the cost of computation is likely to be lower than dealing directly with $\pi_2^N(x_2)$.  Then at time 2 one uses an MCMC kernel
$K_2:\mathbb{R}^{d_x}\times\mathsf{I}_s\times\mathcal{B}(\mathbb{R}^{d_x})\vee\mathcal{I}_s\rightarrow[0,1]$ that admits $\widehat{\pi}_2^N(x_2,i_{1:s})\sigma_{\mathsf{M}_2}^{M_2}(dx_2)\mu_{\mathsf{I}_s}(di_{1:s})$
as its invariant measure.   The way in which this latter kernel is constructed is via a Metropolis-within-Gibbs method
as we now detail.   For any $i_{1:s}\in\mathsf{I}_s$ given,  let $\widehat{K}_2:\mathbb{R}^{d_x}\times\mathsf{I}_s\times\mathcal{B}(\mathbb{R}^{d_x})\rightarrow[0,1]$ be an ergodic Markov-kernel of invariant measure
proportional to $\widehat{\pi}_2^N(x_2,i_{1:s})\sigma_{\mathsf{M}_2}^{M_2}(dx_2)$ (such as in \cite{byrne2013geodesic, graham2017asymptotically, graham2022manifold, zappa2018monte}).  Then $\widehat{K}_2$ is
used to sample from the full-conditional of $X_2$ given $i_{1:s}$.  Let $x_2\in\mathsf{M}_2$ be given and let
 $\check{K}_2:\mathbb{R}^{d_x}\times\mathsf{I}_s\times\mathcal{I}_s\rightarrow[0,1]$ be an ergodic Markov kernel of invariant measure
proportional to $\widehat{\pi}_2^N(x_2,i_{1:s})\mu_{\mathsf{I}_s}(di_{1:s})$; a simple example is a Metropolis-Hastings kernel.  Then we set
$$
K_2\left((x_2,i_{1:s}),(dx_2',di_{1:s}')\right) = \widehat{K}_2\left((x_2,i_{1:s}),dx_2'\right)
\check{K}_2\left((x_2',i_{1:s}),di_{1:s}'\right).
$$
One can produce $N$ samples $(x_2^1,\dots,x_2^N)\in\mathsf{M}_2^N$ and so one can repeat this idea again at time step 3 and later times.

\subsection{Final Algorithm}

For $k\geq 2$, let $\widehat{\pi}_k^N$ be the probability density on $\mathsf{M}_k\times\mathsf{I}_s$ defined by
\begin{equation}\label{eq:smcmc_target}
\widehat{\pi}_k^N(x_k,i_{1:s}) \propto g_k(x_k)\sum_{j\in\mathsf{I}_s}
f_k(x_{k-1}^j,x_k),\qquad
\sigma_{\mathsf{M}_k}^{M_k}\otimes\mu_{\mathsf{I}_s}\text{-a.e.}
\end{equation}
The kernel $K_k$ is constructed via a Metropolis-within-Gibbs sampler that admits $\widehat{\pi}_k^N$ as its invariant density.  Specifically,
\begin{equation}\label{eq:gibbs_kernel}
K_k\left((x_k,i_{1:s}),(dx_k',di_{1:s}')\right) = \widehat{K}_k\left((x_k,i_{1:s}),dx_k'\right)
\check{K}_k\left((x_k',i_{1:s}),di_{1:s}'\right)
\end{equation}
where $\widehat{K}_k$ updates $x_k$ given $i_{1:s}$ via a manifold MCMC kernel (Section~\ref{sec:ex_mcmc}) and $\check{K}_k$ updates $i_{1:s}$ given $x_k$ via a simple Metropolis-Hastings step.  The computational cost per observation time is approximately constant, making it a truly online method.  As noted in the introduction, it circumvents having to design proposals moving from one manifold to the next, which would seem to be a requirement of traditional particle filtering approaches. Our final method is summarised in Algorithm~\ref{alg:smcmc}.

\begin{algorithm}[ht]
\caption{Sequential Manifold MCMC (SMCMC)}
\label{alg:smcmc}
\begin{algorithmic}[1]
\State Choose $(N,s)\in\mathbb{N}\times\{1,\dots,N\}$.
\State \textbf{Initialization.} Run $K_1$ for $N$ steps, obtaining $(X_1^1,\dots,X_1^N)\in\mathsf{M}_1^N$. Set $k=2$.
\While{$k\leq K$}
\State Given samples $(x_{k-1}^1,\dots,x_{k-1}^N)\in\mathsf{M}_{k-1}^N$ from the previous step.
\State Run $K_k$ for $N$ steps using the target density \eqref{eq:smcmc_target}, obtaining $(X_k^1,\dots,X_k^N)\in\mathsf{M}_k^N$.
\State $k\leftarrow k+1$.
\EndWhile
\end{algorithmic}
\end{algorithm}

The SMCMC method allows one to approximate $\pi_k(\varphi)$ for any $k\in\mathbb{N}$ using
$$
\pi_k^N(\varphi) = \frac{1}{N}\sum_{i=1}^N \varphi(x_k^i).
$$
Under appropriate assumptions, it is possible to show convergence of this estimator (in an almost sure sense) to $\pi_k(\varphi)$ as $N\rightarrow\infty$; see \cite{martin2013inference}.

To see why particle filters are ill-suited for this setting, note that for any proposal $q(x_k|x_{k-1}^i,y_k)$ that is absolutely continuous with respect to the Lebesgue measure, the probability of landing on the constraint manifold $\mathsf{M}_k$ is zero. All importance weights therefore vanish, and the filter collapses regardless of the proposal strategy. Our approach avoids this by working directly with the Riemannian measure on $\mathsf{M}_k$.

An important remark is that if manifold MCMC methods are too expensive,  one could relax the problem and consider approximate manifold MCMC as in \cite{escudero2024approximate}.

\subsection{Example of a MCMC Kernel for manifolds}\label{sec:ex_mcmc}

In order to demonstrate that MCMC methods on manifolds are possible,  we briefly review the method of \cite{zappa2018monte} as it will be used in Section \ref{sec:numerics}.  We do not include complete details and these can be found in \cite{zappa2018monte}.

For  simplicity of exposition we shall consider sampling for $k=1$ from $\pi_1(x_1)\sigma_{\mathsf{M}_1}^{M_1}(dx_1)$, but a modification to other contexts is fairly simple to derive.
Let $x\in\mathsf{M}_1$ and denote the tangent space of $x$ as $\mathsf{T}_x$ and $\mathsf{T}_x^{\perp}$ as the
orthogonal space.

Let $\partial c_1(x)^{\top}=QR$ be the thin Q-R decomposition and let $U_x$ be the last $d_x-d_y$ rows of $Q^{\top}$, so that
\begin{equation}\label{eq:tangent_basis}
\mathsf{T}_x=\{U_x^{\top}z:z\in\mathbb{R}^{d_x-d_y}\},\qquad
\mathsf{T}_x^{\perp}=\{(\partial c_1(x))^{\top}a:a\in\mathbb{R}^{d_y}\}.
\end{equation}
Given $v\in\mathsf{T}_x$, the normal component $w\in\mathsf{T}_x^{\perp}$ is the unique solution to $c_1(x+v+w)=0$; the reverse decomposition $(v',w')$ for $x-y$ is obtained analogously from the Q-R decomposition of $\partial c_1(y)^{\top}$.
The acceptance probability for a proposed state $y\in\mathsf{M}_1$ is
\begin{equation}\label{eq:accept_prob}
\alpha(x,y)=\min\left\{1,\frac{\pi_1(y)\,q(v'|x)}{\pi_1(x)\,q(v|x)}\right\},
\quad q(v|x)=\mathcal{N}_{d_x}(0,\rho^2U_x^{\top}U_x)(v).
\end{equation}

We begin at a point $x\in\mathsf{M}_1$ and we now show how the next point in the Markov chain is generated.  Algorithm~\ref{alg:manifold_mcmc} summarises the procedure; full details can be found in \cite{zappa2018monte}.

\begin{algorithm}[ht]
\caption{One Step of the Manifold MCMC Kernel}
\label{alg:manifold_mcmc}
\begin{algorithmic}[1]
\State Compute $U_x$ from the thin Q-R decomposition of $\partial c_1(x)^{\top}$ as in \eqref{eq:tangent_basis}.
\State Sample $Z\sim\mathcal{N}_{d_x-d_y}(0,\mathtt{I})$ and set $v=\rho U_x^{\top}Z$.
\State Solve $c_1(x+v+w)=0$ for $w\in\mathsf{T}_x^{\perp}$; set $y=x+v+w$.
\State Compute $(v',w')$ from the thin Q-R decomposition of $\partial c_1(y)^{\top}$.
\State Accept $y$ with probability $\alpha(x,y)$ from \eqref{eq:accept_prob}; otherwise stay at $x$.
\end{algorithmic}
\end{algorithm}

There is an important point which is omitted in the above steps.  The root-finding procedure used to compute $w$ must be deterministic and parameterised by a pre-specified tolerance and maximum number of iterations.  After obtaining $(v',w')$ from $y$, one must check whether applying the same root-finding method to $y$ with $v'$ recovers $x$; if it does not, the move is immediately rejected.  Work  \cite{zappa2018monte} provides a complete explanation.

The computational cost of the SMCMC method is dominated by the evaluation of the auxiliary target density $\widehat{\pi}_k^N$ in Algorithm~\ref{alg:smcmc}. Each evaluation requires the sum over $s$ ancestor samples $\sum_{j\in\mathsf{I}_s}f_k(x_{k-1}^j,x_k)$ together with the Jacobian determinant $g_k(x_k)$. If one used all $N$ previous samples ($s=N$), the cost to generate $N$ new samples would be $\mathcal{O}(N^2)$ per time step, as noted in Section~\ref{sec:method}. The introduction of the conditioning set reduces this to $\mathcal{O}(N s)$. In practice $s$ is chosen as a fixed integer much smaller than $N$, although ideally it would grow with $d_x$ to control the Monte Carlo error of the approximation to the predictive density.

The per-proposal cost of the manifold MCMC kernel (Algorithm~\ref{alg:manifold_mcmc}) consists of the thin Q-R decomposition of the $d_x\times d_y$ matrix $\partial c_k(x)^\top$, costing $\mathcal{O}(d_x d_y^2)$, and the solution of the constraint equation $c_k(x+v+w)=0$ for $w\in\mathsf{T}_x^\perp$ via Newton's method. Each Newton iteration requires an evaluation of $c_k$ and its Jacobian $\partial c_k$, together with the solution of a $d_y\times d_y$ linear system, giving a total cost of $\mathcal{O}(d_y^2 d_x + d_y^3)$ per proposal when Newton is needed. For linear constraints the Newton solve is unnecessary and the Q-R decomposition can be cached, reducing the cost to $\mathcal{O}(d_x d_y)$. In problems where $d_y$ is proportional to $d_x$ (for instance when many components are observed), the cubic term $d_y^3$ may become significant.

End-to-end, the SMCMC algorithm applied to $n$ observation times with $N$ samples and a conditioning set of size $s$ costs $\mathcal{O}\big(n\,N\,(s\,T(d_x) + d_y^2 d_x + d_y^3)\big)$, where $T(d_x)$ is the cost of a single transition density evaluation. For the models considered in Section~\ref{sec:numerics}, $T(d_x)$ is either $\mathcal{O}(1)$ or $\mathcal{O}(d_x)$, so the overall cost is dominated by the $\mathcal{O}(n\,N\,s\,d_x)$ term. For models requiring a matrix factorisation (e.g. a Cholesky decomposition of a dense $d_x\times d_x$ covariance), $T(d_x)$ could be $\mathcal{O}(d_x^3)$, but such cases are not considered in this work. The per-step cost does not increase with $n$, making the method suitable for online filtering. The practical cost and stability of the kernel also depend on the root-finding tolerance and the maximum number of Newton iterations; for the examples in this paper these settings were adequate, with linear constraints detected automatically and solved without iteration.

\subsection{Convergence of Low Noise to Degenerate Noise Case}\label{sec:conv_res}

In this section we consider a special case of the filtering problem and the convergence,  in an appropriate sense,  of the low noise sequential MCMC method; we now give the details.  We restrict ourselves to the model that for each $k\in\mathbb{N}$:
$$
Y_k = A_kX_k + \Delta^{1/2}\epsilon_k
$$
where $\{A_k\}_{k\geq 1}$ is a sequence of $d_y\times d_x$ real matrices with full row rank,  $d_x>d_y$,  $\epsilon_k\stackrel{\text{i.i.d.}}{\sim}\mathcal{N}_{d_y}(0,\mathtt{I}_{d_y})$ and $\Delta>0$ for the low noise filter, whereas $\Delta=0$ for the corresponding degenerate noise filter. For the dynamics $\{X_k\}_{k\geq 0}$ we do not make any assumptions except that the Lebesgue transition density $f_k(x_{k-1},x_k)$ can be evaluated pointwise. In the sequel we will sometimes refer to the low noise filter and other times to the degenerate noise one. The distinction will be obvious from the context and the particular notation used in each setting.

As the sequence of manifolds $\widetilde{\mathsf{M}}_k$ (resp.~$\mathsf{M}_k$) are determined via linear constraints,  it is possible in this setting to specify Lebesgue densities of the filters on $d_x-$dimensional (resp.~$(d_x-d_y)-$dimensional) spaces.
At time $k$ we use $\widetilde{Z}_k\in\mathbb{R}^{d_x}$ to denote the unknown random variable associated to the filter at time $k$ of the low noise problem,  and use the notation $\widetilde{Z}_k=(Z_k,\overline{Z}_k)^{\top}$ where
$Z_k\in\mathbb{R}^{d_x-d_y}$ and $\overline{Z}_k\in\mathbb{R}^{d_y}$.  We can write the Lebesgue density
$\widetilde{\pi}_k^{\lambda}(\widetilde{z}_k)$ in the following manner.  Let $A_k^{\Delta}:=[A_k,\Delta^{1/2}\mathtt{I}_{d_y}]$ be a $d_y\times(d_x+d_y)$ matrix,  define $\text{ker}(A_k^{\Delta}):=\{x\in\mathbb{R}^{d_x+d_y}:
A_k^{\Delta}x = 0\}$,  let $V_k^{\Delta}$ be a $(d_x+d_y)\times d_x$
matrix whose columns are an
 orthonormal basis of $\text{ker}(A_k^{\Delta})$ and $\widetilde{z}_k^{\Delta}$ be a $(d_x+d_y)-$dimensional vector that solves $Y_k=A_k^{\Delta}\widetilde{z}_k^{\Delta}$.
To compute $\widetilde{z}_k^{\Delta}$,  let  $z_k^{\star}$ be any $d_x-$dimensional vector that solves $Y_k=A_kz_k^{\star}$, then we set $\widetilde{z}_k^{\Delta}=(z_k^{\star},0)^{\top}$.
Finally for any $z\in\mathbb{R}^{d_x}$ set
$$
u_k^{\Delta}(z) = \widetilde{z}_k^{\Delta} + V_k^{\Delta} z
$$
and we will use the convention that $u_k^{\Delta}(z) = (u_k^{\Delta}(z,x),u_k^{\Delta}(z,\epsilon))^{\top}$.
$u_k^{\Delta}(z,x)\in\mathbb{R}^{d_x}$,  $u_k^{\Delta}(z,\epsilon)\in\mathbb{R}^{d_y}$.
Note also that $Y_k=A_k^{\Delta}u_k^{\Delta}(z)$.
Then we have that
$$
\widetilde{\pi}_1^{\lambda}(\widetilde{z}_1) \propto p_{\mathtt{N}}(u_1^{\Delta}(\widetilde{z}_1,\epsilon))
f_1(x_0,u_1^{\Delta}(\widetilde{z}_1,x))
$$
and for any $k\geq 2$
$$
\widetilde{\pi}_k^{\lambda}(\widetilde{z}_k) \propto p_{\mathtt{N}}(u_k^{\Delta}(\widetilde{z}_k,\epsilon))
\int_{\mathbb{R}^{d_x}}
f_k(u_{k-1}^{\Delta}(\widetilde{z}_{k-1},x), u_k^{\Delta}(\widetilde{z}_k,x))
\widetilde{\pi}_{k-1}^{\lambda}(\widetilde{z}_{k-1})\lambda_{d_x}(d\widetilde{z}_{k-1})).
$$
For the case of the degenerate noise,  we can write the Lebesgue density
$\pi_k^{\lambda}(z_k)$ in a similar manner,  which is added for completeness.
Let  $\text{ker}(A_k) =\{x\in\mathbb{R}^{d_x}:
A_kx = 0\}$,  $V_k^{\star}$ be a $d_x\times (d_x-d_y)$
matrix whose columns are an
 orthonormal basis of $\text{ker}(A_k)$ and $z_k^{\star}$ is as used above.  For any $z\in\mathbb{R}^{d_x-d_y}$ set
$$
u_k^{\star}(z) = z_k^{\star} + V_k^{\star} z
$$
and we note that $Y_k=A_ku_k^{\star}(z)$.
Then we have that
$$
\pi_1^{\lambda}(z_1) \propto
f_1(x_0,u_1^{\star}(z_1))
$$
and for any $k\geq 2$
$$
\pi_k^{\lambda}(z_k) \propto
\int_{\mathbb{R}^{d_x}}
f_k(u_{k-1}^{\star}(z_{k-1}),u_k^{\star}(z_k))
\pi_{k-1}^{\lambda}(z_{k-1})\lambda_{d_x-d_y}(dz_{k-1})).
$$

For the MCMC kernels required by the SMCMC method,  we will use in the low-noise case a sequence of random walk Metropolis kernels with Gaussian proposals in dimension $d_x$.  We focus on the $\mathcal{O}(N^2)$ implementation (with $s=N$) for simplicity although relaxing this is a matter of changing the notations,  rather than mathematical
complexity.  The MCMC kernel,  $K_k^{\Delta}:\mathbb{R}^{d_x}\times\mathcal{B}(\mathbb{R}^{d_x})\rightarrow[0,1]$,  can be described at any time $k\geq 2$,  for
$(\widetilde{z}_{k-1}^1,\dots,\widetilde{z}_{k-1}^N)\in\mathbb{R}^{Nd_x}$ given
\begin{eqnarray*}
K_k^{\Delta}(\widetilde{z},d\widetilde{z}') & = &  \min\left\{1,
\frac{
p_{\mathtt{N}}(u_k^{\Delta}(\widetilde{z}',\epsilon))
\tfrac{1}{N}
\sum_{i=1}^N f_k(u_{k-1}^{\Delta}(\widetilde{z}_{k-1}^i,x), u_k^{\Delta}(\widetilde{z}',x))
}{
p_{\mathtt{N}}(u_k^{\Delta}(\widetilde{z},\epsilon))
\tfrac{1}{N}
\sum_{i=1}^N f_k(u_{k-1}^{\Delta}(\widetilde{z}_{k-1}^i,x), u_k^{\Delta}(\widetilde{z},x))
}
\right\} \widetilde{q}_k(\widetilde{z},\widetilde{z}') \lambda_{d_x}(d\widetilde{z}') + \\ & &
\delta_{\{\widetilde{z}\}}(d\widetilde{z}')\Bigg[1-
\int_{\mathbb{R}^{d_x}}
\min\left\{1,
\frac{
p_{\mathtt{N}}(u_k^{\Delta}(\widetilde{z}',\epsilon))
\tfrac{1}{N}
\sum_{i=1}^N f_k(u_{k-1}^{\Delta}(\widetilde{z}_{k-1}^i,x), u_k^{\Delta}(\widetilde{z}',x))
}{
p_{\mathtt{N}}(u_k^{\Delta}(\widetilde{z},\epsilon))
\tfrac{1}{N}
\sum_{i=1}^N f_k(u_{k-1}^{\Delta}(\widetilde{z}_{k-1}^i,x), u_k^{\Delta}(\widetilde{z},x))
}
\right\} \times \\ & & \widetilde{q}_k(\widetilde{z},\widetilde{z}') \lambda_{d_x}(d\widetilde{z}')
\Bigg]
\end{eqnarray*}
where $\widetilde{q}_k(\widetilde{z},\widetilde{z}')$ is the Gaussian proposal density used at time $k$
and $\delta_{\{z\}}(dz')$ is the Dirac measure on the set $\{z\}$. The case $k=1$ is similar and is not written as it should be clear from the context.  For $(\widetilde{z},\widetilde{z}_{k-1}^1,\dots,\widetilde{z}_{k-1}^N)\in\mathbb{R}^{(N+1)d_x}$
given and any $k\geq 2$ we write the conditional expectation,  associated to $K_k^{\Delta}$, of any $\varphi:\mathbb{R}^{d_x}\rightarrow\mathbb{R}$ that is bounded and measurable (write the collection of such functions $\mathtt{B}_b(\mathbb{R}^{d_x})$) as
$$
\mathbb{E}_k^{\Delta}\left[\varphi(\widetilde{Z}')\Big|\widetilde{z},\widetilde{z}_{k-1}^1,\dots,\widetilde{z}_{k-1}^N\right] = \int_{\mathbb{R}^{d_x}} \varphi(\widetilde{z}')K_k^{\Delta}(\widetilde{z},d\widetilde{z}').
$$
If $k=1$ we simply write the expectation associated to  $K_1^{\Delta}$
$$
\mathbb{E}_1^{\Delta}\left[\varphi(\widetilde{Z}')\Big|\widetilde{z}\right].
$$

 In the degenerate case we can consider
a sequence of random walk Metropolis kernels with Gaussian proposals in dimension $d_x-d_y$.  Again,  the MCMC kernel,  $K_k^{\star}:\mathbb{R}^{d_x-d_y}\times\mathcal{B}(\mathbb{R}^{d_x-d_y})\rightarrow[0,1]$,  can be described at any time $k\geq 2$,  for
$(z_{k-1}^1,\dots,z_{k-1}^1)\in\mathbb{R}^{N(d_x-d_y)}$ given
\begin{eqnarray*}
K_k^{\star}(z,dz') & = &  \min\left\{1,
\frac{
\tfrac{1}{N}
\sum_{i=1}^N f_k(u_{k-1}^{\star}(z_{k-1}^i), u_k^{\star}(z'))
}{
\tfrac{1}{N}
\sum_{i=1}^N f_k(u_{k-1}^{\star}(z_{k-1}^i), u_k^{\star}(\widetilde{z}))
}
\right\} q_k(z,z') \lambda_{d_x-d_y}(dz') + \\ & &
\delta_{\{z\}}(dz')\Bigg[1-
\int_{\mathbb{R}^{d_x-d_y}}
\min\left\{1,
\frac{
\tfrac{1}{N}
\sum_{i=1}^N f_k(u_{k-1}^{\star}(z_{k-1}^i), u_k^{\star}(z'))
}{
\tfrac{1}{N}
\sum_{i=1}^N f_k(u_{k-1}^{\star}(z_{k-1}^i), u_k^{\star}(z))
}
\right\}
q_k(z,z') \lambda_{d_x-d_y}(dz')
\Bigg]
\end{eqnarray*}
where $q_k(z,z')$ is the Gaussian proposal density used at time $k$.
Again,  we do not explicitly write the case $k=1$ as it should be fairly clear to the reader.

We are now ready to state our main result whose proof is in Appendix \ref{app:app2}.
Recall that we use the notation $\widetilde{z}=(z,\overline{z})^{\top}$.
\begin{prop}\label{prop:prop2}
For any $(\varphi,\widetilde{z})\in\mathtt{B}_b(\mathbb{R}^{d_x-d_y})\times\mathbb{R}^{d_x}$ we have that
\begin{eqnarray*}
\lim_{\Delta\downarrow 0} \mathbb{E}_1^{\Delta}\left[\varphi(Z')\Big|\widetilde{z}\right] & = &
\int_{\mathbb{R}^{d_x}}\varphi(z')
\min\left\{1,
\frac{
p_{\mathtt{N}}(\bar{z}')
 f_1(x_0, u_1^{\star}(z'))
}{
p_{\mathtt{N}}(\bar{z})
f_1(x_0, u_1^{\star}(z))
}
\right\}
\widetilde{q}_1(\widetilde{z},\widetilde{z}')\lambda_{d_x}(d\widetilde{z}')
+ \\ & &
\varphi(z)\left[
1 - \int_{\mathbb{R}^{d_x}}
\min\left\{1,
\frac{
p_{\mathtt{N}}(\bar{z}')
 f_1(x_0, u_1^{\star}(z'))
}{
p_{\mathtt{N}}(\bar{z})
f_1(x_0, u_1^{\star}(z))
}
\right\}
\widetilde{q}_1(\widetilde{z},\widetilde{z}')\lambda_{d_x}(d\widetilde{z}')
\right].
\end{eqnarray*}
In addition,  for any $(\varphi,k,N,\widetilde{z})\in\mathtt{B}_b(\mathbb{R}^{d_x-d_y})\times\{2,3,\dots\}\times\mathbb{N}\times\mathbb{R}^{d_x}$
and $(\widetilde{z}_{k-1}^1,\dots,\widetilde{z}_{k-1}^N)\in\mathbb{R}^{Nd_x}$
we have that
$$
\lim_{\Delta\downarrow 0}
\mathbb{E}_k^{\Delta}\left[\varphi(Z')\Big|\widetilde{z},\widetilde{z}_{k-1}^1,\dots,\widetilde{z}_{k-1}^N\right]  =
$$
$$
\int_{\mathbb{R}^{d_x}}\varphi(z')
\min\left\{1,
\frac{
p_{\mathtt{N}}(\bar{z}')
\tfrac{1}{N}
\sum_{i=1}^N f_k(u_{k-1}^{\star}(z_{k-1}^i), u_k^{\star}(z'))
}{
p_{\mathtt{N}}(\bar{z})
\tfrac{1}{N}
\sum_{i=1}^N f_k(u_{k-1}^{\star}(z_{k-1}^i), u_k^{\star}(z))
}
\right\}
\widetilde{q}_k(\widetilde{z},\widetilde{z}')\lambda_{d_x}(d\widetilde{z}')
+
$$
$$
\varphi(z)\left[
1 - \int_{\mathbb{R}^{d_x}}
\min\left\{1,
\frac{
p_{\mathtt{N}}(\bar{z}')
\tfrac{1}{N}
\sum_{i=1}^N f_k(u_{k-1}^{\star}(z_{k-1}^i), u_k^{\star}(z'))
}{
p_{\mathtt{N}}(\bar{z})
\tfrac{1}{N}
\sum_{i=1}^N f_k(u_{k-1}^{\star}(z_{k-1}^i), u_k^{\star}(z))
}
\right\}
\widetilde{q}_k(\widetilde{z},\widetilde{z}')\lambda_{d_x}(d\widetilde{z}')
\right].
$$
\end{prop}

The result says,  in our context,  as we send the noise to zero,  performing SMCMC on the low noise case converges to an SMCMC which, marginally,  has an invariant measure that is the degenerate noise case.   Indeed if one chose
$$
\widetilde{q}_k(\widetilde{z},\widetilde{z}') = q_k(z,z')p_{\mathtt{N}}(\bar{z})
$$
then,  marginally,  one would be using the kernel $K^{\star}_k$ in the limit as $\Delta\downarrow 0$. This result has important implications from the point of view of robustness of computational efficiency with decreasing $\Delta$ close to zero.
If in the degenerate case $K_k^{\star}$ mixes well then one should expect good mixing when using SMCMC in the low noise case.  As remarked previously,  generic SMCMC without the methodology proposed in this paper can struggle to be efficient in the low noise case even when using sophisticated state of the art MCMC algorithms due to the conditional likelihood of $Y_k$ given $X_k$ being too informative.

We note that Proposition~\ref{prop:prop2} is restricted to linear observation models.  Extending the convergence analysis to nonlinear observation functions is an open problem.  Furthermore, the result establishes only pointwise convergence of the MCMC kernels, not convergence of the filtering distributions themselves.  Consistency of the SMCMC estimator $\pi_k^N(\varphi)$ follows from the general SMCMC theory of \cite{martin2013inference} under standard assumptions on the Markov kernel and the state-space model.

\section{Numerical Results}\label{sec:numerics}

We illustrate the performance of the proposed SMCMC method on five numerical experiments covering linear and nonlinear observation operators, low- and high-dimensional state spaces, and the transition from low to degenerate observation noise.  For brevity we consider only the degenerate noise case, which is the most challenging for standard filtering algorithms.

Across all experiments we use the manifold MCMC kernel of Section~\ref{sec:ex_mcmc} with a target acceptance rate of $0.234$ \cite{gelman1997weak,beskos2018asymptotic}, achieved via a Robbins-Monro adaptive step-size scheme during a warmup phase of $500$ iterations (step-size adaptation rate $0.01$).  The initial proposal step-size is set to $\rho=0.25$ for the linear and nonlinear Gaussian models, $\rho=0.35$ for the FHN model, and $\rho=0.1$ for the KS model.  The root-finding parameters are a Newton tolerance of $10^{-12}$, a reversibility check tolerance of $10^{-8}$, a maximum of $50$ Newton iterations, and automatic detection of linear constraints. The percentage of rejections due to projection errors has consistently stayed below $0.1\%$ for all examples. All experiments use a fixed random seed ($42$). The effective sample size (ESS) is computed using Geyer's initial positive sequence estimator \cite{geyer1992practical}; for multi-dimensional states the ESS is averaged over components.  The normalised mean-squared error (NMSE) is computed relative to the Kalman filter (where available) or the simulated true state.  For each experiment we run $N=10^4$ MCMC iterations ($N=4\times10^4$ for the high-dimensional KS example) with a conditioning set of size $s\in\{10,20,50\}$.  Comparisons are made against the Kalman filter (linear-Gaussian models), joint manifold MCMC using the same Zappa et al.~\cite{zappa2018monte} kernel, and, where available, manifold Hamiltonian Monte Carlo via the \texttt{mici} library \cite{graham2022manifold}.

\subsection{Linear Gaussian Model: Robustness to Observation Noise}

We begin with a problem that bridges the low-noise and degenerate-noise regimes.  Consider the model
\begin{align*}
    Y_k &= \mathbf{1}^\top X_k + \sqrt{\Delta}\,\epsilon_k,\qquad \epsilon_k\sim\mathcal{N}(0,1),\\
    X_k &= \alpha X_{k-1} + \sigma W_k,\qquad W_k\sim\mathcal{N}(0,\mathtt{I}_{d_x}),
\end{align*}
with $d_x=20$, $n=30$, $\alpha=0.6$, $\sigma=0.5$, and $\mathbf{1}^\top$ the row vector of all ones.  For $\Delta=0$ the observation is degenerate and the filter lives on the hyperplane $\{x:\mathbf{1}^\top x=y_k\}$.  For $\Delta>0$ we embed the noise as an auxiliary variable $\epsilon_k$ and work on the augmented manifold $\{(x,\epsilon):\mathbf{1}^\top x+\sqrt{\Delta}\,\epsilon = y_k\}\subset\mathbb{R}^{d_x+1}$, so that a single algorithm handles all noise levels seamlessly.  This connects directly to the convergence result of Proposition~\ref{prop:prop2}.

\begin{figure}[htbp]
    \centering
    \includegraphics[width=\linewidth]{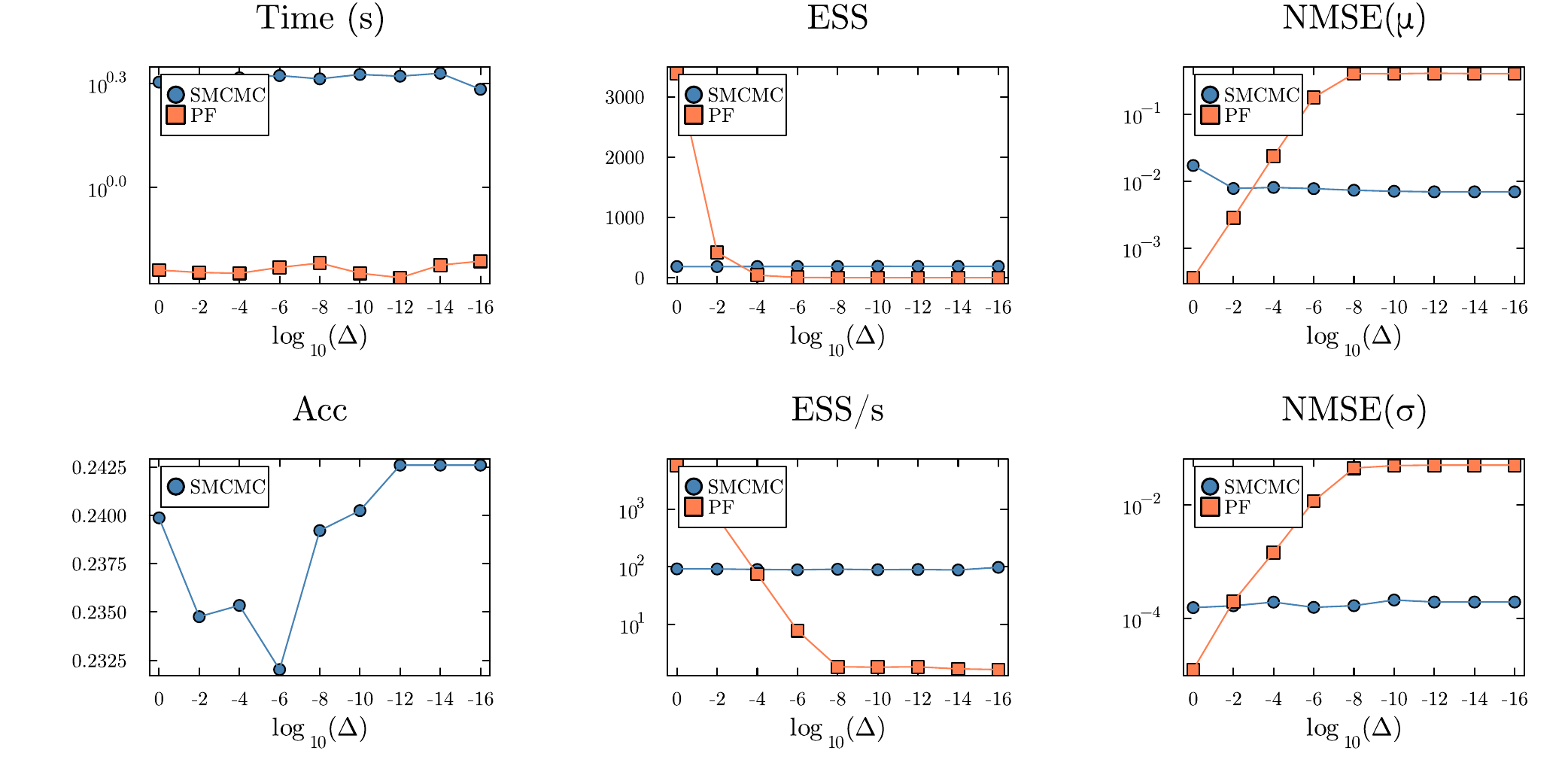}
    \caption{Linear Gaussian model: sweep over observation noise $\Delta$.  Six panels show runtime, acceptance rate, ESS, ESS/s, NMSE of the mean, and NMSE of the standard deviation against $\log_{10}\Delta$.  SMCMC (blue) maintains stable performance while the bootstrap PF (orange) collapses as $\Delta\to0$.}
    \label{fig:ex01}
\end{figure}

We run SMCMC ($N=10^4$, $s=50$) and a bootstrap particle filter ($10^4$ particles) across nine noise levels $\Delta\in\{10^0,10^{-2},\dots,10^{-16}\}$.  The results are summarized in Figure~\ref{fig:ex01}.  At $\Delta=1$ both methods perform well.  As $\Delta$ drops below $10^{-4}$ the PF collapses: ESS falls from $419$ ($\Delta=10^{-2}$) to $42$ ($10^{-4}$) and then to $1$ for $\Delta\le10^{-8}$, while the NMSE of the mean inflates to about $0.4$.  SMCMC is essentially unaffected: the acceptance rate stays near $0.234$, the ESS is stable at $185$-$190$, and both NMSE metrics remain stable across the entire range.  These results illustrate that SMCMC remains robust across the entire range of noise levels, consistent with the convergence analysis of Proposition~\ref{prop:prop2}.

\subsection{Linear Gaussian Model: Hyperplane Constraint}

We now turn to the fully degenerate observation model
\begin{align*}
    Y_k &= \mathbf{1}^\top X_k,\\
    X_k &= \alpha X_{k-1} + \sigma W_k,\qquad W_k\sim\mathcal{N}(0,\mathtt{I}_{d_x}),
\end{align*}
with $d_x=20$, $n=30$, $\alpha=0.6$, $\sigma=0.5$.  The constraint $\mathbf{1}^\top x_k = y_k$ defines a hyperplane; because the constraint is linear, the tangent-space basis $U_x$ is constant and the Q-R decomposition can be cached, so the per-proposal cost reduces to $\mathcal{O}(d_x d_y)$.

\begin{figure}[htbp]
    \centering
    \includegraphics[width=\linewidth]{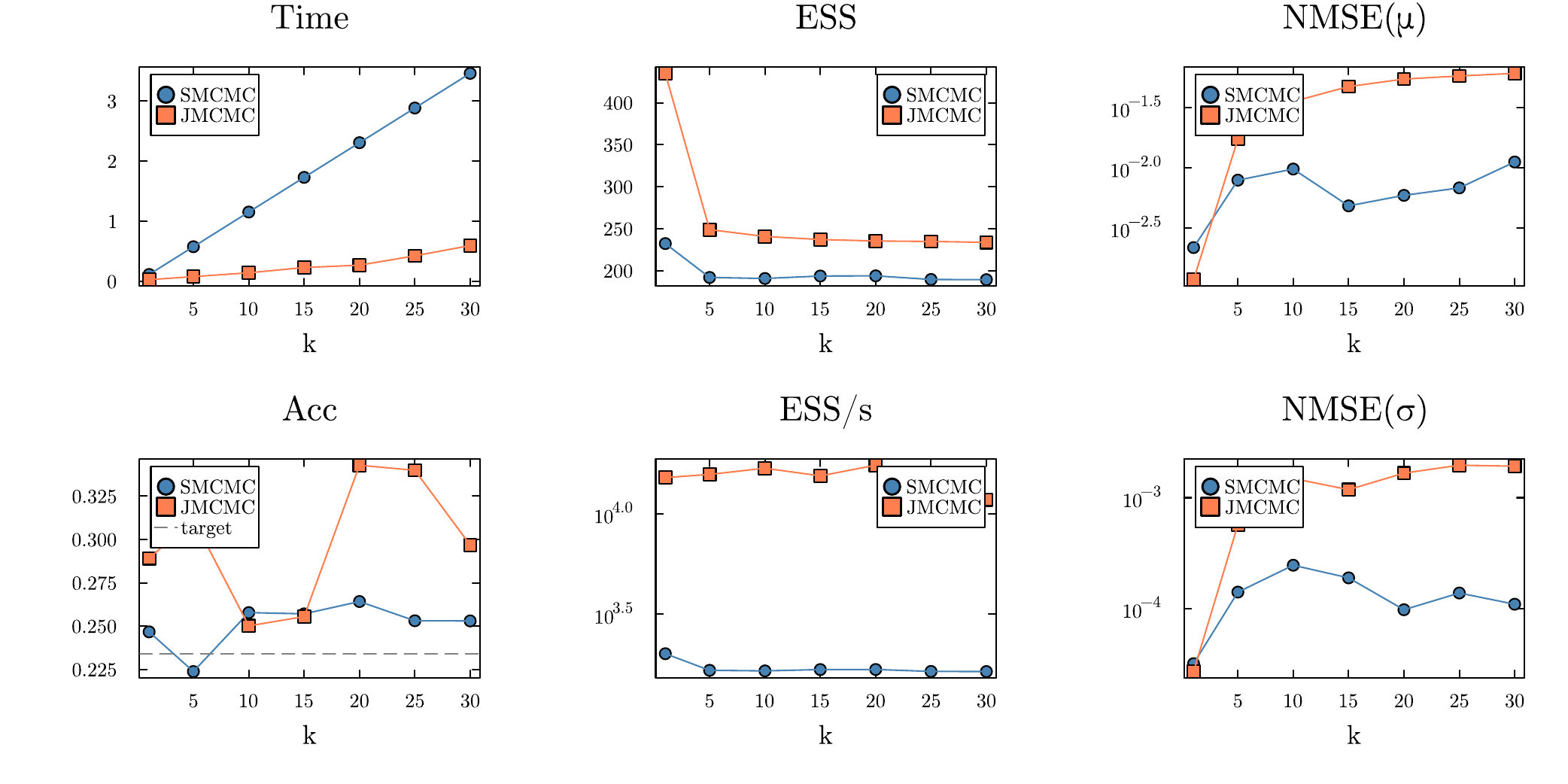}
    \caption{Linear Gaussian model with hyperplane constraint.  SMCMC (blue) vs.\ JMCMC (orange).  The six panels show wall-clock time, acceptance rate (grey dashed: target $0.234$), ESS, ESS/s, NMSE of the mean, and NMSE of the standard deviation (both against the KF).}
    \label{fig:ex11}
\end{figure}

We compare SMCMC ($N=10^4$, $s=50$) against the Kalman filter (KF) and against joint manifold MCMC (JMCMC), which samples the product manifold $\pmb{\mathsf{M}}_k$ and extracts the marginal at time $k$.  Results are shown in Figure~\ref{fig:ex11}.  The JMCMC ESS is higher at early time steps ($435$ at $k=1$) but degrades steadily as $k$ grows, whereas the SMCMC ESS stays in the range $190$-$230$ regardless of $k$.  The ESS/s of SMCMC is stable at roughly $1650$ for $k\ge5$, while the JMCMC ESS/s drops from around $15{,}000$ to about $11{,}500$ as the product-manifold dimension grows.  For this example, joint MCMC achieves higher ESS, although its computational cost increases with the time horizon. On the contrary, the per-step cost of SMCMC is constant, making it more suitable for longer data sequences and more complex problems.

\begin{figure}[htbp]
    \centering
    \includegraphics[width=\linewidth]{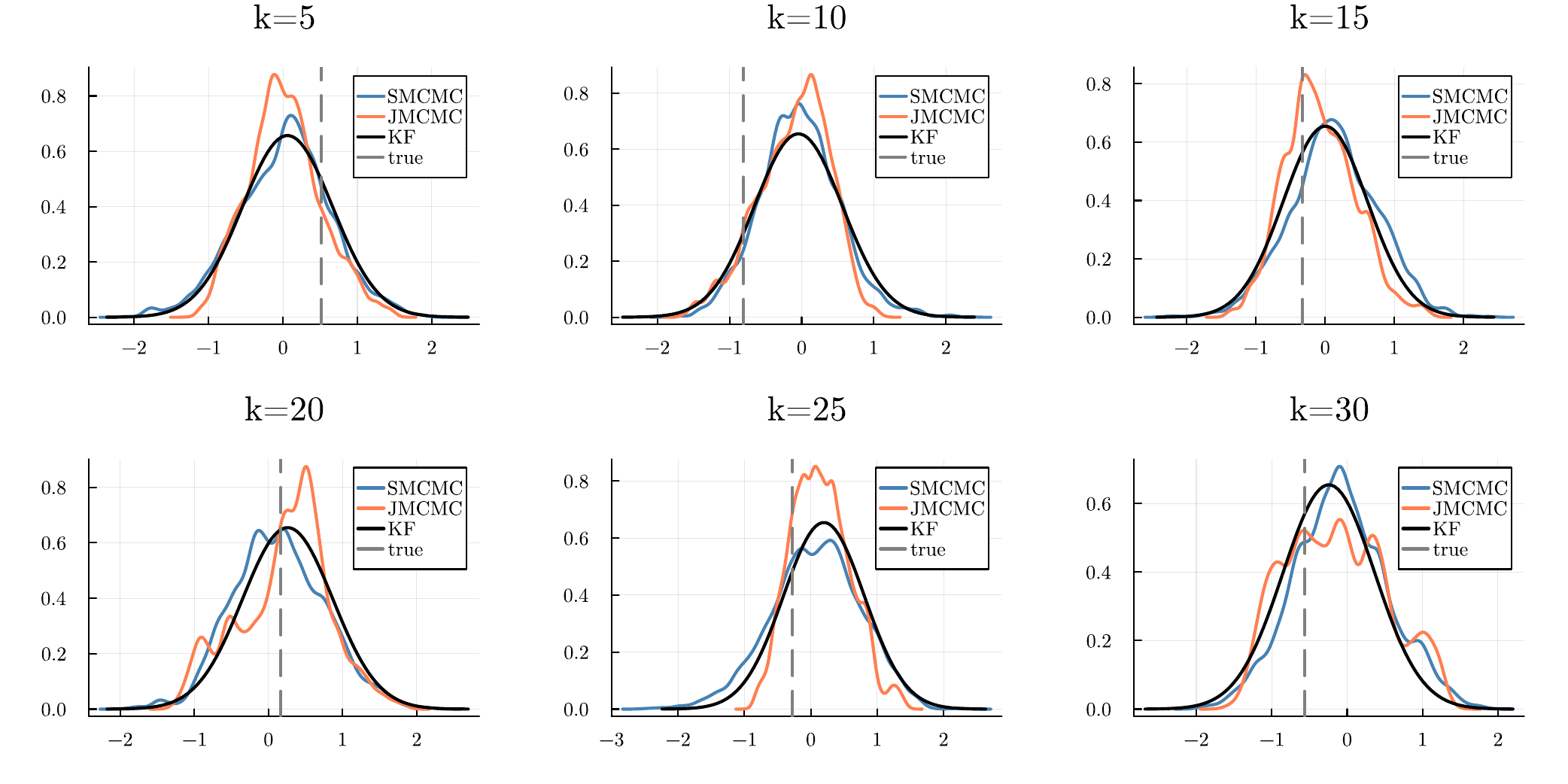}
    \caption{Linear Gaussian model (hyperplane).  Marginal densities of $X_{k,2}$ at six time steps.  SMCMC (blue), KF (black), true value (dashed).}
    \label{fig:ex12}
\end{figure}

Marginal posterior densities of the second component $X_{k,2}$ at $k\in\{5,10,15,20,25,30\}$ are shown in Figure~\ref{fig:ex12}.  The SMCMC estimates (blue) closely match the exact KF densities (black). The JMCMC (orange) looks underdispersed, indicating that more samples are needed. This is unsurprising, the joint target is $600$-dimensional.

\subsubsection*{Choice of the conditioning set size $s$}

The auxiliary target \eqref{eq:smcmc_target} replaces the full $N$-sample predictive sum with a sum over $s$ randomly chosen ancestors, reducing the per-iteration cost from $\mathcal{O}(N^2)$ to $\mathcal{O}(Ns)$.  To study the effect of $s$, we run SMCMC with $s\in\{1,5,10,20,50,100,200,500,1000\}$, keeping $N=10^4$ and the target acceptance rate fixed to~$0.234$.

\begin{figure}[htbp]
    \centering
    \includegraphics[width=\linewidth]{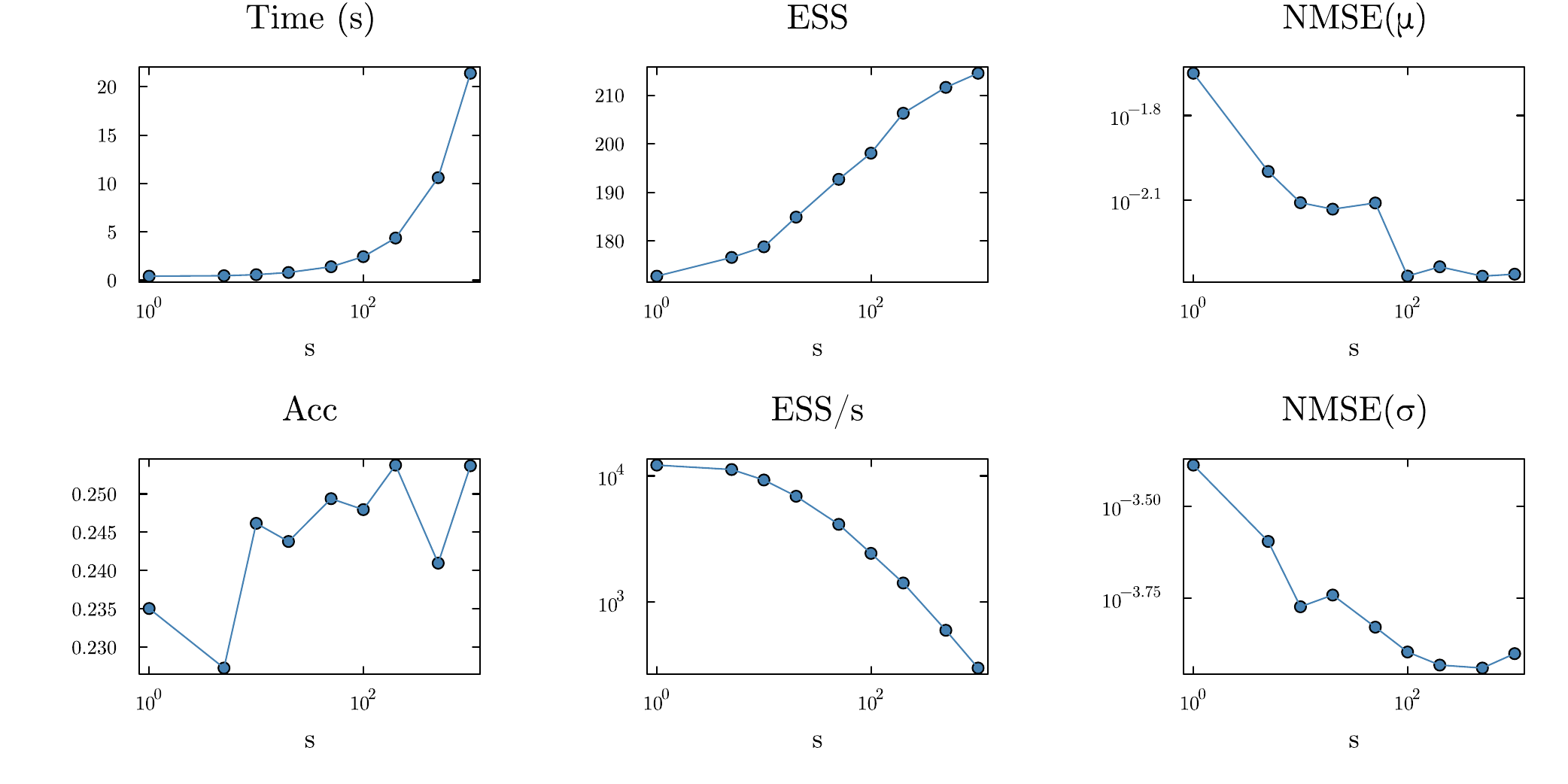}
    \caption{Linear Gaussian model: effect of the conditioning set size $s$.  ESS (top left), ESS/s (top right), NMSE of the mean (bottom left), NMSE of the standard deviation (bottom centre) and total runtime (bottom right).}
    \label{fig:ex13}
\end{figure}

{\small
\begin{table}[t]
\footnotesize
\centering
\caption{Linear Gaussian model -- choice of conditioning set size $s$}
\label{tab:ex1_s_sweep}
\begin{tabular}{lrrrrrrrrr}
\toprule
 & $s=1$ & $s=5$ & $s=10$ & $s=20$ & $s=50$ & $s=100$ & $s=200$ & $s=500$ & $s=1000$ \\
\midrule
Time & 0.43 & 0.47 & 0.58 & 0.80 & 1.40 & 2.44 & 4.36 & 10.60 & 21.39 \\
Acc & 0.235 & 0.227 & 0.246 & 0.244 & 0.249 & 0.248 & 0.254 & 0.241 & 0.254 \\
ESS & 173 & 177 & 179 & 185 & 193 & 198 & 206 & 212 & 215 \\
ESS/s & 12165 & 11224 & 9272 & 6896 & 4132 & 2435 & 1418 & 599 & 301 \\
NMSE($\mu$) & 0.02237 & 0.01005 & 0.00778 & 0.00739 & 0.00777 & 0.00428 & 0.00462 & 0.00428 & 0.00435 \\
NMSE($\sigma$) & 0.00041 & 0.00025 & 0.00017 & 0.00018 & 0.00015 & 0.00013 & 0.00012 & 0.00011 & 0.00013 \\
\bottomrule
\end{tabular}
\end{table}

}

Results are shown in Figure~\ref{fig:ex13} and Table~\ref{tab:ex1_s_sweep}.  With $s=1$ the ESS is $173$ and the NMSE of the mean is approximately $0.022$.  Increasing $s$ to $50$ raises the ESS to $193$ and reduces the NMSE to $0.008$, while the runtime grows only moderately (from $0.43$ to $1.40$~seconds).  Beyond $s=50$ the gains plateau: at $s=1000$ the ESS reaches $215$ but the runtime jumps to $21.5$~seconds, reducing the ESS/s by a factor of $40$.  A value of $s=50$ offers a good balance between accuracy and cost, and we adopt it for the following experiments unless otherwise noted.

\subsection{Gaussian Model with Nonlinear Observations}

We now consider a nonlinear observation constraint:
\begin{align*}
    Y_k &= \|X_k\|^2,\\
    X_k &= \alpha X_{k-1} + \sigma W_k,\qquad W_k\sim\mathcal{N}(0,\mathtt{I}_{d_x}),
\end{align*}
so that the manifold is the sphere $\{x:\|x\|^2 = y_k\}$.  We set $d_x=20$, $n=30$, $\alpha=0.6$, $\sigma=0.5$.  The Jacobian $2x^\top$ has full rank away from the origin, and the nonlinearity requires the Newton retraction step in Algorithm~\ref{alg:manifold_mcmc}.

\begin{figure}[htbp]
    \centering
    \includegraphics[width=\linewidth]{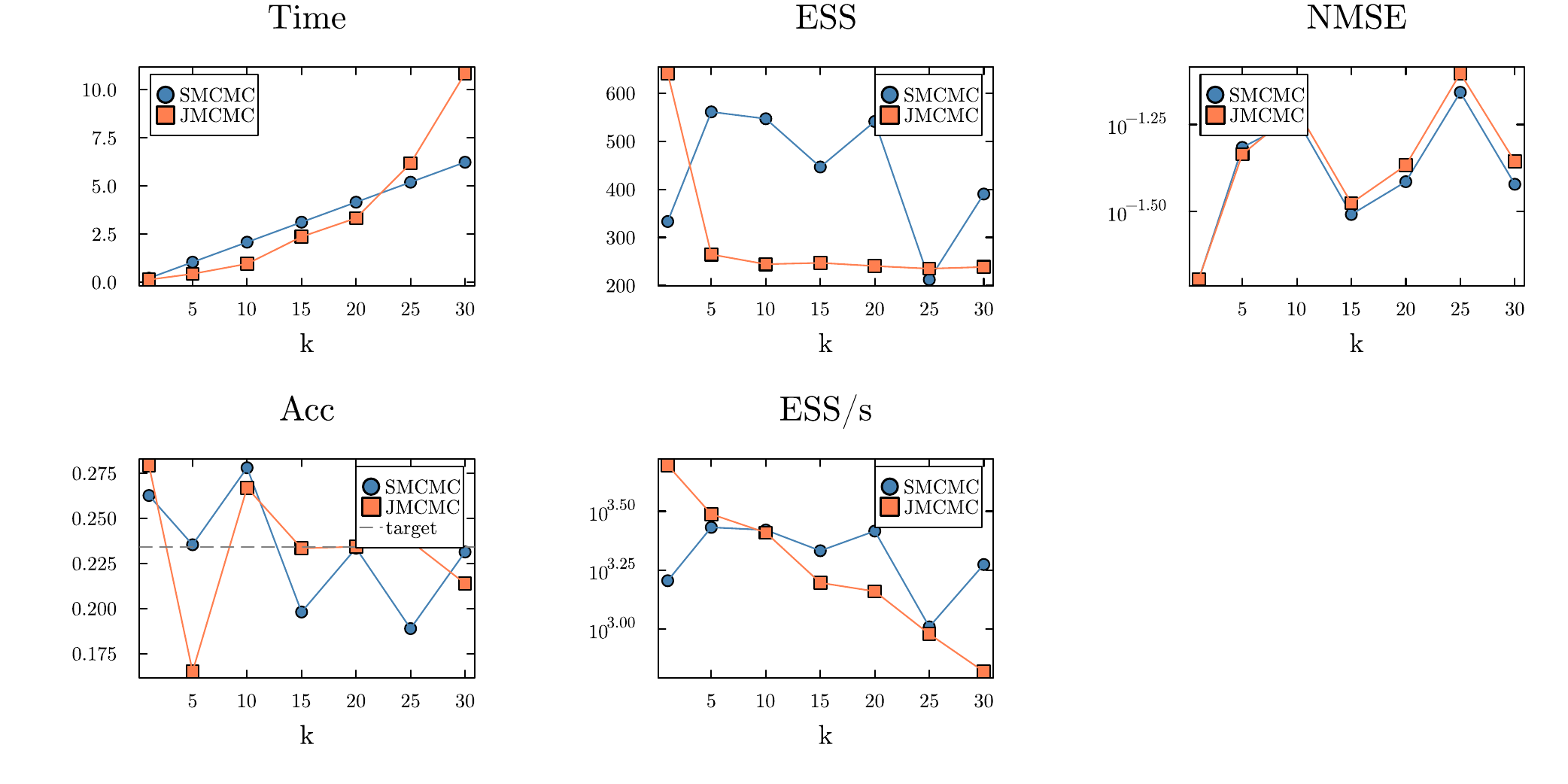}
    \caption{Gaussian model with sphere constraint.  SMCMC (blue) vs.\ JMCMC (orange).  Panels show time, acceptance rate, ESS, ESS/s (log scale), and NMSE against the true state.}
    \label{fig:ex21}
\end{figure}

The conditioning set size is $s=50$, as suggested by a sweep over $s$ (similar to LGM study). Results are shown in Figure~\ref{fig:ex21}.  The SMCMC acceptance rate is close to the target $0.234$ at most time steps, and the ESS ranges from $210$ to $560$, comparable to or higher than the JMCMC ESS at the same $k$.  The ESS/s of SMCMC is around $1900$ at $k=30$, nearly three times larger than the JMCMC value of $650$, confirming the advantage of the sequential approach on curved manifolds. Notice the changepoint at $k=10$, when the ESS/s advantage shifts from JMCMC to SMCMC.

\begin{figure}[htbp]
    \centering
    \includegraphics[width=\linewidth]{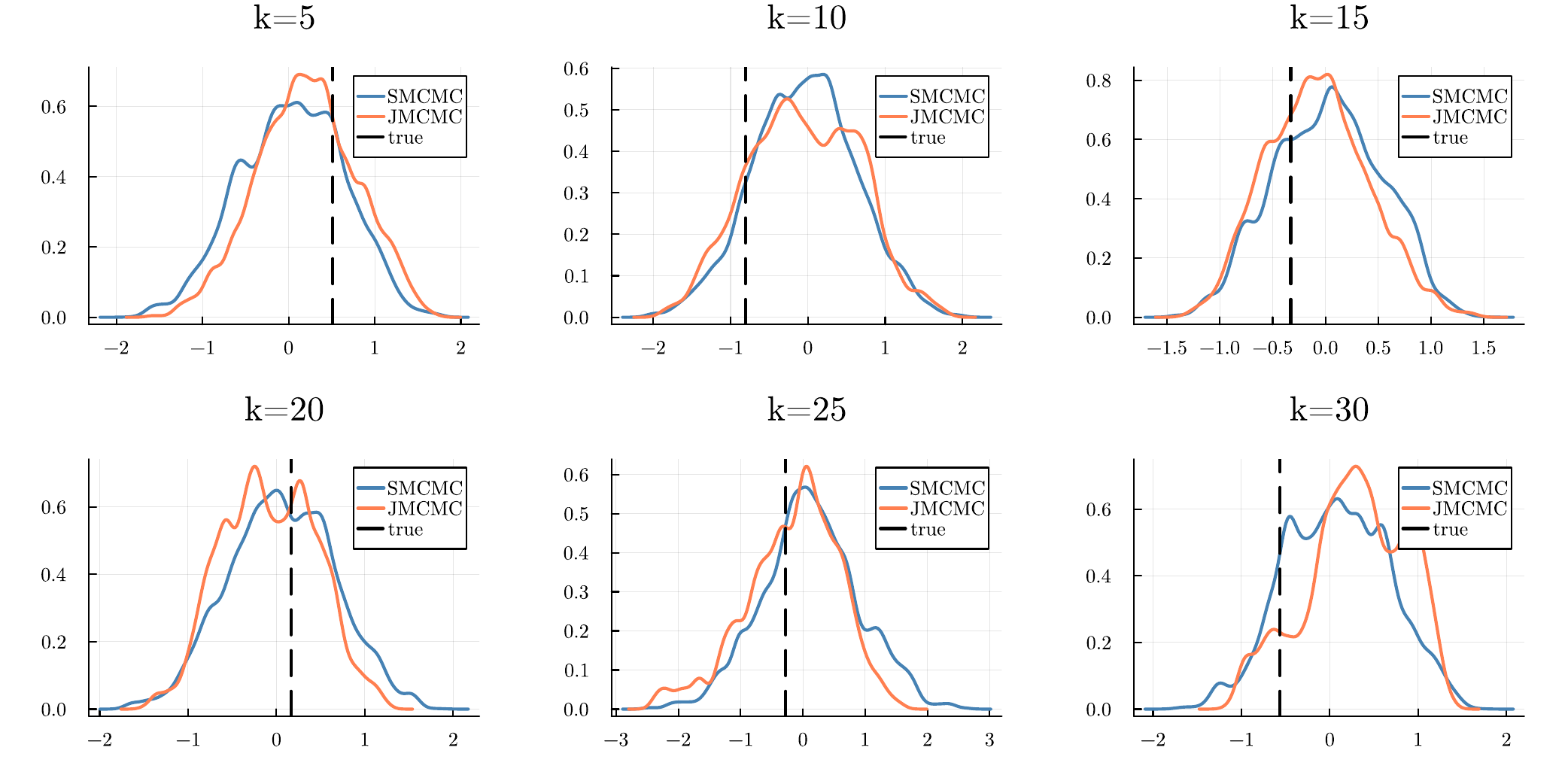}
    \caption{Gaussian model with sphere constraint.  Marginal densities of $X_{k,2}$ at $k=5,10,15,20,25,30$.  SMCMC (blue) and JMCMC (orange) produce consistent estimates.}
    \label{fig:ex22}
\end{figure}

Marginal posterior densities of the second component at several time steps (Figure~\ref{fig:ex22}) show good agreement between SMCMC and JMCMC. The sphere constraint is genuinely nonlinear, and the stable performance of SMCMC suggests that the Newton retraction does not present practical difficulties for these examples.

\subsection{FitzHugh-Nagumo Model}

\begin{figure}[htbp]
    \centering
    \includegraphics[width=\linewidth]{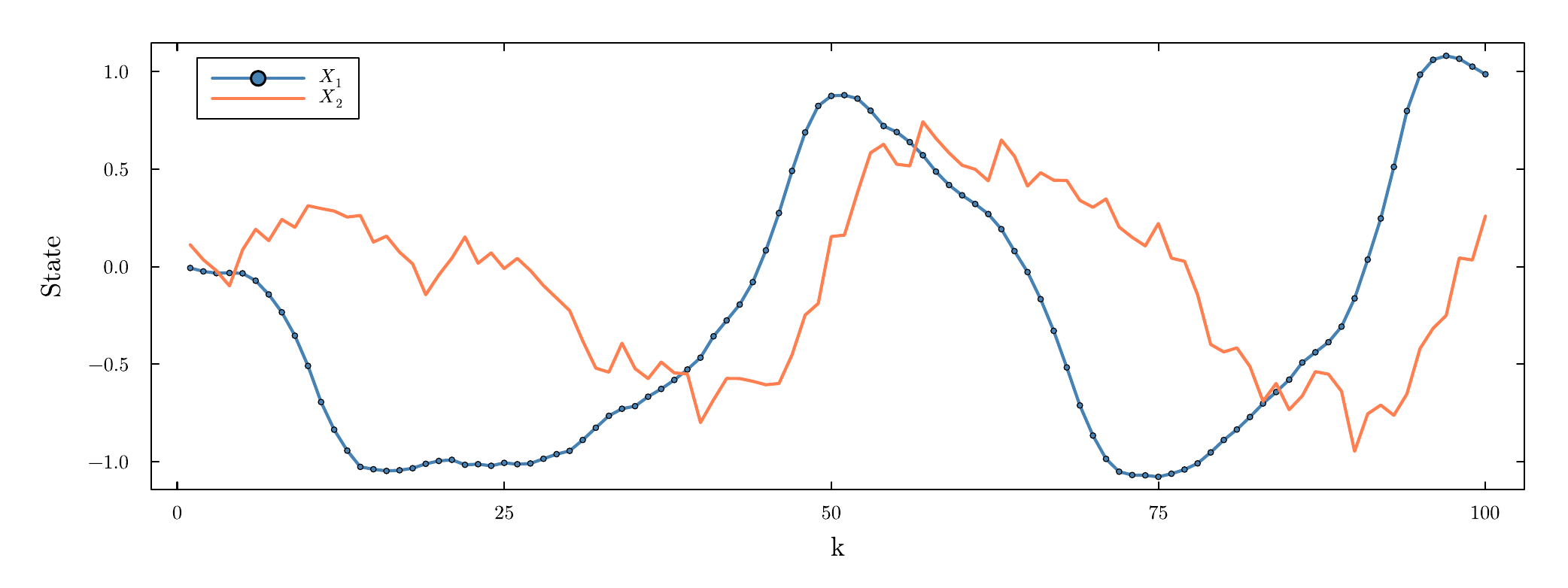}
    \caption{FHN model.  True state trajectory over time: observed component $X_1$ (black) and hidden recovery variable $X_2$ (orange).}
    \label{fig:ex33}
\end{figure}

We now apply SMCMC to the FitzHugh-Nagumo (FHN) model, a $2$-dimensional hypoelliptic diffusion that models neural spiking:
\begin{align*}
    dX_1 &= \epsilon^{-1}(X_1 - X_1^3 - X_2)\,dt,\\
    dX_2 &= (\gamma X_1 - X_2 + \beta)\,dt + \sigma\,dW_t,
\end{align*}
with $\sigma=0.5$, $\epsilon=0.2$, $\gamma=1.5$, $\beta=0.5$.  The SDE is discretised via a strong order $1.5$ Taylor scheme with step $\delta=0.05$ and $n=100$ time steps.  Only the first component is observed ($Y_k = X_1(t_k)$), so the filtering inference focuses on the unobserved recovery variable $X_2$.  Figure~\ref{fig:ex33} shows the simulated state trajectory.

\begin{figure}[htbp]
    \centering
    \includegraphics[width=\linewidth]{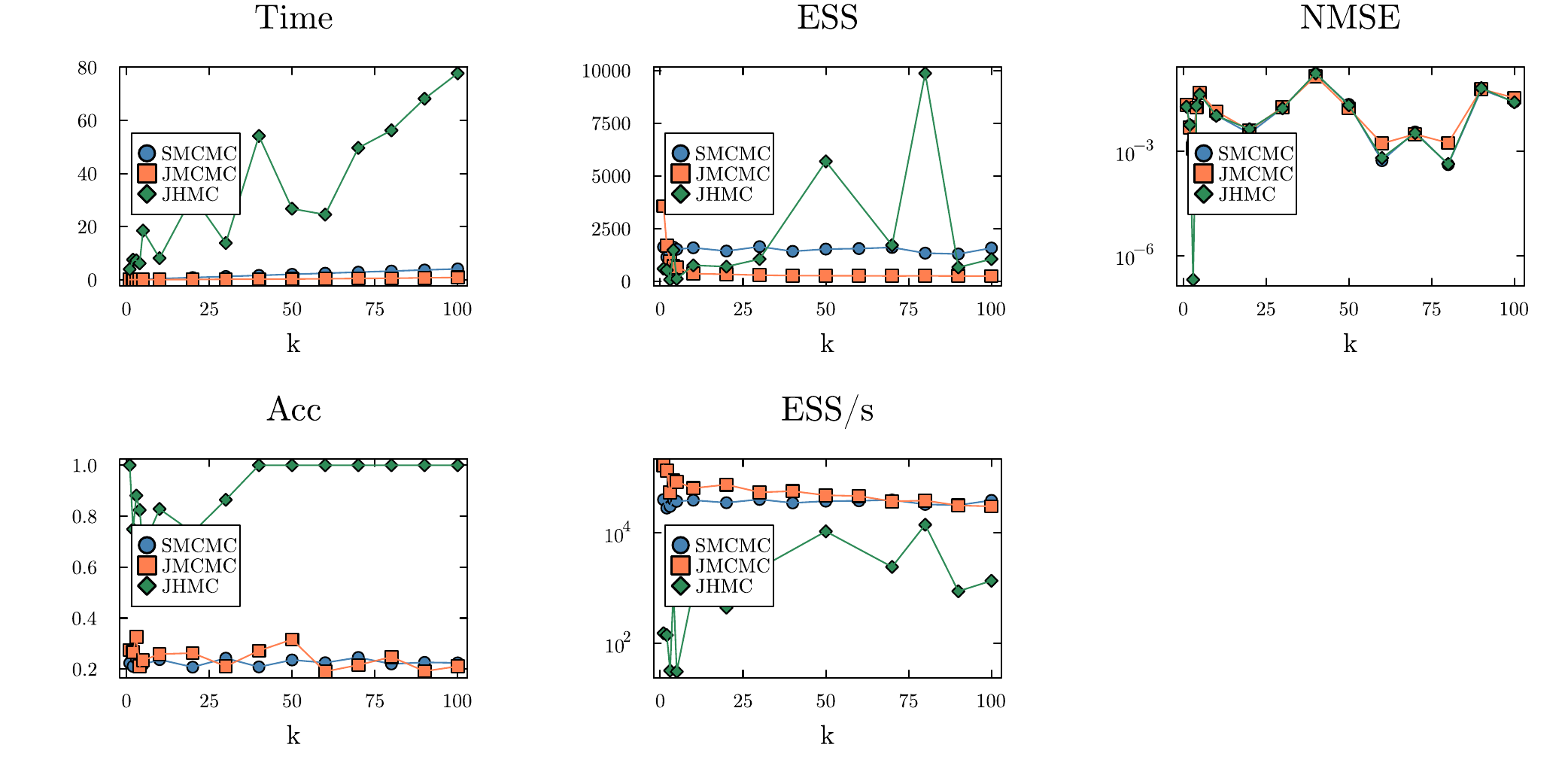}
    \caption{FHN model.  SMCMC (blue), JMCMC (orange), JHMC (green).  Panels show time, acceptance rate, ESS, ESS/s (log scale), and NMSE against the true $X_2$.}
    \label{fig:ex31}
\end{figure}

We run SMCMC ($N=10^4$, $s=20$) and compare with JMCMC and with Hamiltonian Monte Carlo (JHMC) on a subset of time indices (750 iterations with 250 warm-up using the \texttt{mici} library). The conditioning set $s=20$ was chosen as a moderate value, although for this low-dimensional problem ($d_x=2$) the choice of $s$ has a limited effect on ESS and even smaller values give comparable results. Results are in Figure~\ref{fig:ex31}.  The SMCMC ESS for $X_2$ stays in $1150$-$1650$ with no systematic decay across all $100$ steps.  The JMCMC ESS starts at $3500$ at $k=1$ but drops to $250$ at $k=100$, reflecting the growing product-manifold dimension.  The ESS/s of SMCMC (${\sim}30{,}000$-$40{,}000$) exceeds that of JMCMC at large $k$, and achieves higher ESS/s than \texttt{mici}. We note that we faced issues with initialization, projection and adaptation of \texttt{mici}, and thus, many $k$ runs have acceptance rate $1.0$. We interpret the HMC results with caution and emphasize the practical challenges of using more sophisticated kernels. The NMSE values are comparable across all three methods.

\begin{figure}[htbp]
    \centering
    \includegraphics[width=\linewidth]{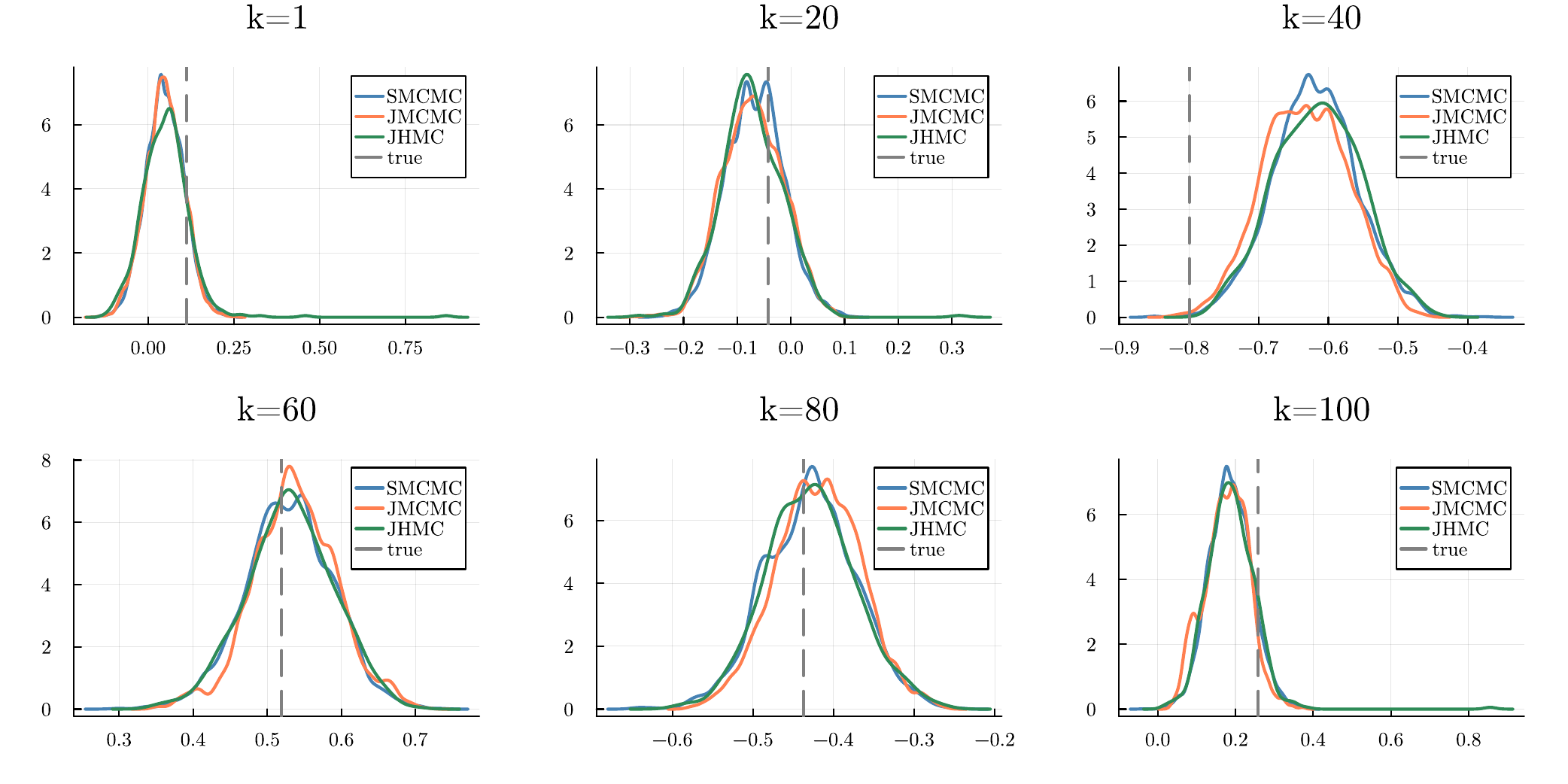}
    \caption{FHN model.  Marginal densities of $X_2$ at $k\in\{1,20,40,60,80,100\}$.  SMCMC (blue), JMCMC (orange), JHMC (green).}
    \label{fig:ex32}
\end{figure}

Figure~\ref{fig:ex32} shows the marginal posterior densities of $X_2$ at six time steps.  The SMCMC densities are in close agreement with JMCMC and JHMC.

\subsection{Kuramoto-Sivashinsky Equation}

\begin{figure}[htbp]
    \centering
    \includegraphics[width=\linewidth]{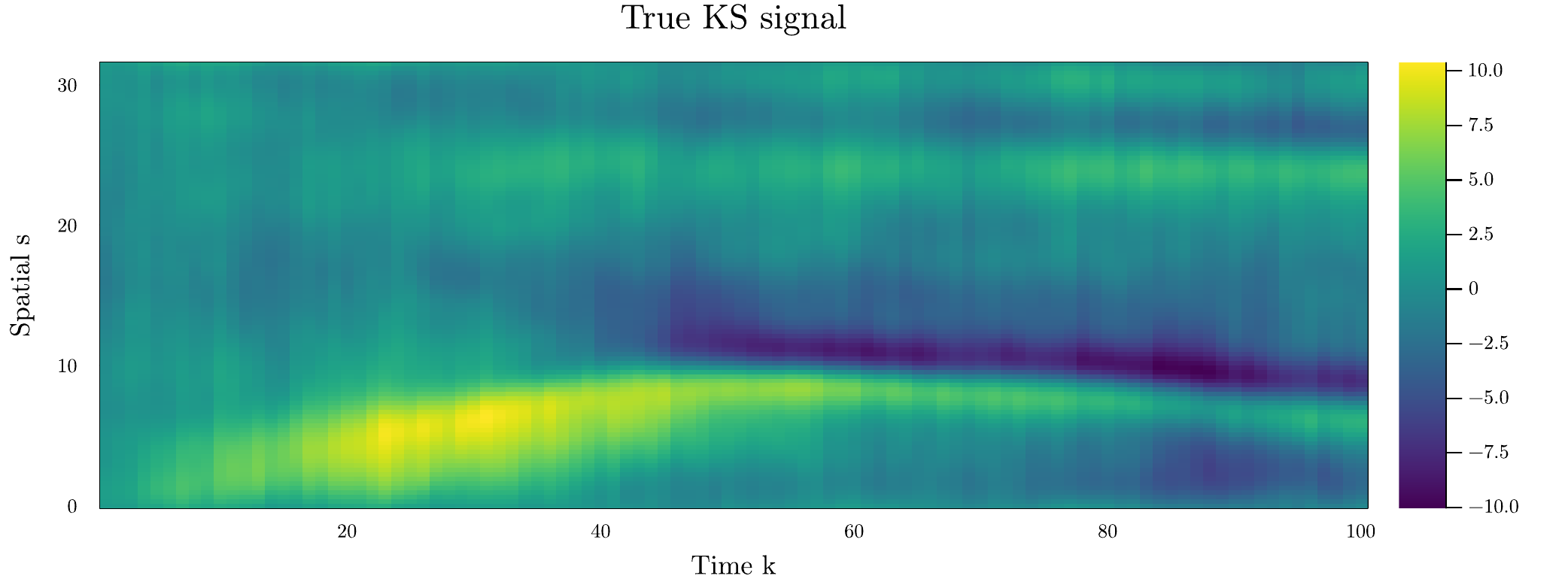}
    \caption{KS model.  Heatmap of the true state $X(s,t)$ over space and time.}
    \label{fig:ex43}
\end{figure}

Our final example is a high-dimensional filtering problem arising from the Kuramoto-Sivashinsky (KS) SPDE,
\begin{equation*}
    \frac{\partial X}{\partial t} + \frac{\partial^2 X}{\partial s^2} + \frac{\partial^4 X}{\partial s^4} + X\frac{\partial X}{\partial s} + \gamma X = \frac{d\mathcal{W}}{dt},
\end{equation*}
posed on $[0,10\pi]$ with periodic boundary conditions and damping $\gamma=0.01$.  The equation is discretised using $d_x=100$ spatial grid points and an implicit-explicit Euler scheme with time step $\delta_t=\delta_s^2/2$, giving
\begin{equation*}
    AX_k = X_{k-1} - X_{k-1}\odot B X_{k-1} + C\nu_k,\qquad \nu_k\sim\mathcal{N}(0,\mathtt{I}_{d_x}),
\end{equation*}
where $A$, $B$ encode the finite-difference stencils for the linear and nonlinear terms, and $C$ is the Cholesky factor of a Mat\'ern$(1/2)$ spatial covariance (range $10$, variance $4$).  The observation is $Y_k = H X_k$ with $H$ selecting every $10$th component, so $d_y=10$.  Figure~\ref{fig:ex43} shows the simulated state field.

To improve mixing in $100$ dimensions we precondition via $V_k = P^{-1}X_k$, where $P$ is the Cholesky factor of $\operatorname{Cov}(X_k\mid Y_k,X_{k-1})$ under a hypothetical noisy model $Y_k\mid X_k\sim\mathcal{N}(HX_k,\sigma_y^2\mathtt{I})$.  This is equivalent to using the Riemannian metric $M_k=\Sigma_k^{-1}$, and it substantially improves mixing and ESS in this example.

\begin{figure}[htbp]
    \centering
    \includegraphics[width=\linewidth]{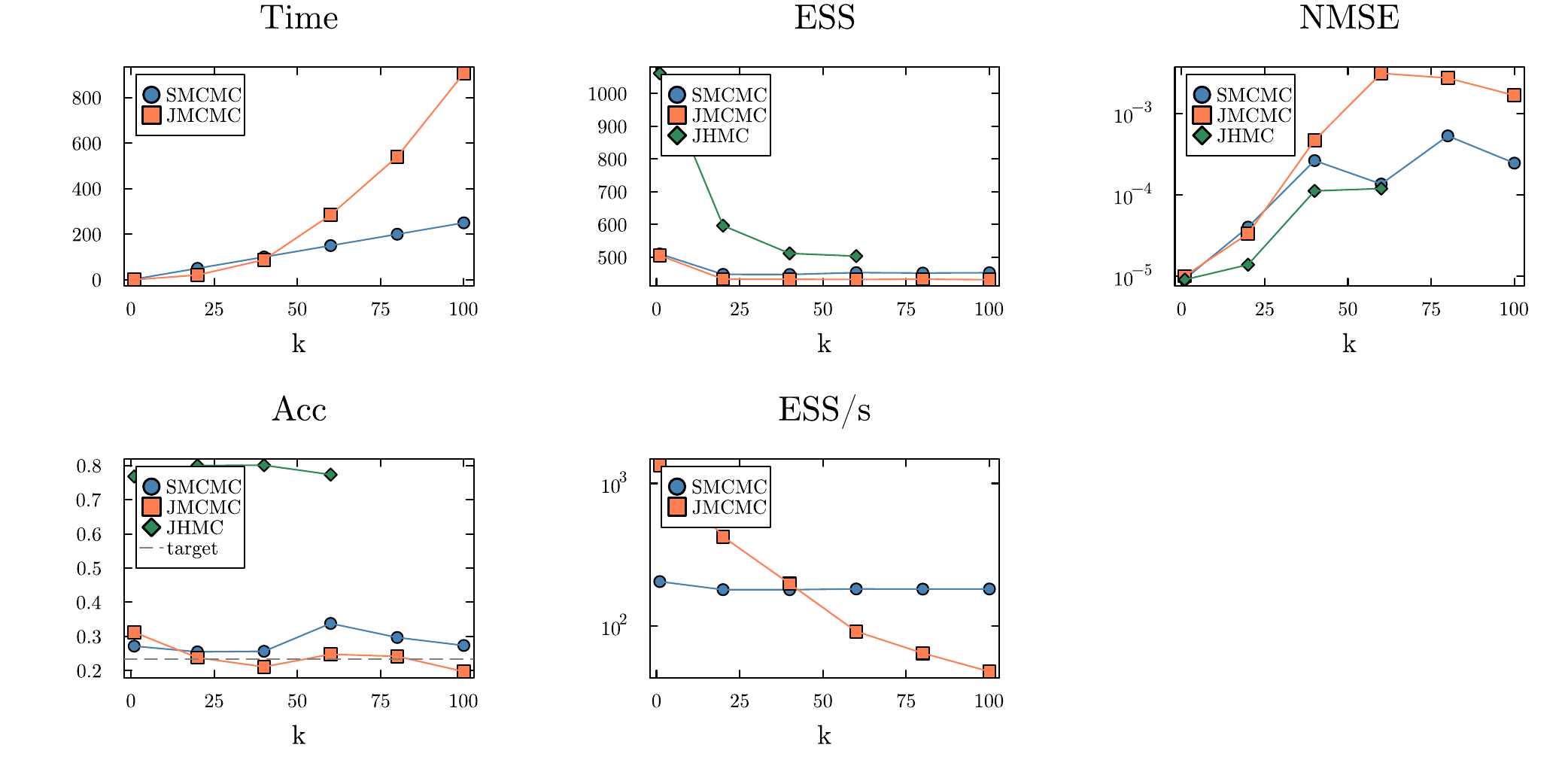}
    \caption{KS model.  SMCMC (blue), JMCMC (orange).  Panels show time, acceptance rate, ESS (averaged over unobserved components), ESS/s (log scale), and NMSE against the true state.}
    \label{fig:ex41}
\end{figure}

We run SMCMC ($N=4\times10^4$, $s=10$ chosen via sweep) and compare with JMCMC at $k\in\{1,20,40,60,80,100\}$.  Results are in Figure~\ref{fig:ex41}.  The SMCMC acceptance rate stays near $0.234$ and the mean ESS over the $90$ unobserved dimensions is stable at $450-510$ across all $k$.  The JMCMC ESS/s starts at $1350$ at $k=1$ but drops to $48$ at $k=100$ as the joint space grows; the SMCMC ESS/s stays constant at approximately $180$. At $k=40$ we observe the takeover. We note that JHMC runs were too slow (up to 20 hours) and we have not included the timings and ESS/s metric in the plot, because they are orders of magnitude worse and worsen the visualisation. The NMSE values are of order $10^{-5}$-$10^{-4}$ for SMCMC (slightly higher for JMCMC at large $k$, and slightly lower for JHMC at a few small $k$ where we could run the latter), confirming that SMCMC does not sacrifice much accuracy.

\begin{figure}[htbp]
    \centering
    \includegraphics[width=\linewidth]{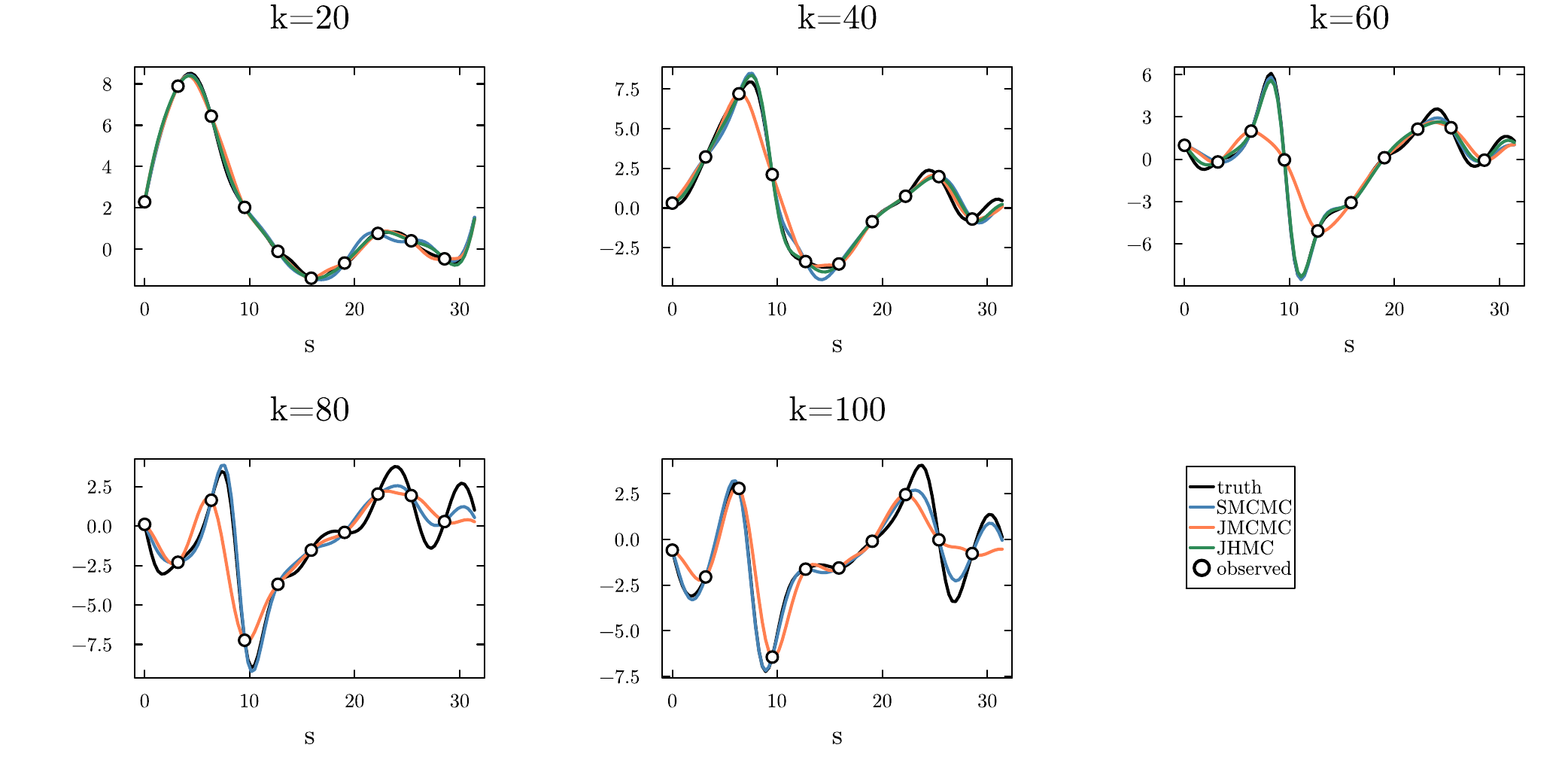}
    \caption{KS model.  Mean field estimates at $k\in\{20,40,60,80,100\}$: true state (black), SMCMC (blue), JMCMC (orange), JHMC (green).}
    \label{fig:ex44}
\end{figure}

The field estimates in Figure~\ref{fig:ex44} show that SMCMC (blue) and JHMC (green, where available) both accurately track the true state across all spatial locations.  JMCMC (orange), by contrast, deviates notably from the truth at several time steps, particularly at larger $k$.  This is consistent with the ESS degradation observed for JMCMC: as the joint manifold dimension grows, the sampler struggles to explore the posterior efficiently, leading to poorer mean estimates.

\section*{Conclusions}

We have developed a sequential MCMC framework for filtering with degenerate or low observation noise by formulating the problem on a sequence of Riemannian manifolds defined by the observation constraints.  The method uses a general-purpose manifold MCMC kernel requiring no custom proposal design and maintains a constant per-step cost, unlike joint MCMC.  The methodology is illustrated on five challenging examples, illustrating robustness to observation noise, stable ESS across time, and accurate tracking of the true state in high dimensions via preconditioning. The present theoretical analysis is restricted to linear observation operators in the convergence result of \ref{sec:conv_res}. Extending these results to nonlinear observations and establishing convergence of the filtering distributions themselves remain important directions for future work.

\subsubsection*{Acknowledgements}
AZ \& AJ were supported by CUHK-SZ start up funding.

\section*{Data availability}

All the data used and analyzed in Section \ref{sec:numerics} was simulated using the code available at \url{https://github.com/abylayzhumekenov/smcmc_noise}, along with all the implementation and figure/table generation.

\bibliographystyle{siamplain}
\bibliography{references}

\appendix

\section{Proofs}\label{app:proofs}

\subsection{Proof of Proposition \ref{prop:prop1}}\label{app:app1}

\begin{proof}
For the case of the filter at time 1,  we can just apply \cite[Proposition 2(b)]{diaconis2013sampling} along with the equivalence
of the Riemannian and Hausdorff measure (e.g.~\cite[Lemma C.2.]{graham2022manifold}).  For $k\geq 2$ we recall that the filter is the time $k$ marginal of the smoother at time $k$.  We will find the density of the smoother $\pmb{\pi}_k(x_{1:k})$ w.r.t.~$\sigma_{\pmb{\mathsf{M}}_k}^{\mathbf{M}_k}$.  Applying again \cite[Proposition 2(b)]{diaconis2013sampling} along with the equivalence of the Riemannian and Hausdorff measure
$$
\pmb{\pi}_k(x_{1:k}) \propto
 \mathrm{det}\left(\partial \mathbf{c}_k(x_{1:k})\mathbf{M}_k^{-1}\partial \mathbf{c}_k(x_k)^{\top}\right)^{-1/2}
\prod_{s=1}^k f_s(x_{s-1},x_s).
$$
Then due to the block structure of $\partial \mathbf{c}_k$ and $\mathbf{M}_k$ we have
$$
 \mathrm{det}\left(\partial \mathbf{c}_k(x_{1:k})\mathbf{M}_k^{-1}\partial \mathbf{c}_k(x_k)^{\top}\right)^{-1/2}
= \prod_{s=1}^k g_s(x_s)
$$
and by standard properties of the Riemannian measure
$\mu_{\pmb{M}_k}^{\mathbf{M}_k}=\bigotimes_{s=1}^k \sigma_{\mathsf{M}_s}^{M_s}$.

As a result of the above arguments,  we have that
\begin{align*}
\pi_k(x_k) & =
\frac
{
g_k(x_k) \int_{\pmb{\mathsf{M}}_{k-1} }
f(x_{k-1},x_k) \prod_{s=1}^{k-1} g_s(x_s)  f_s(x_{s-1},x_s)
\bigotimes_{s=1}^{k-1} \sigma_{\mathsf{M}_s}^{M_s}(dx_s)
}
{
\int_{\pmb{\mathsf{M}}_{k}}
 \prod_{s=1}^{k} g_s(x_s)  f_s(x_{s-1},x_s)
\bigotimes_{s=1}^{k-1} \sigma_{\mathsf{M}_s}^{M_s}(dx_s)
}\\
& =
g_k(x_k) \int_{\mathsf{M}_{k-1}}
f(x_{k-1},x_k)
\left\{ \int_{\pmb{\mathsf{M}}_{k-2} }
\prod_{s=1}^{k-1} g_s(x_s)  f_s(x_{s-1},x_s)
\bigotimes_{s=1}^{k-2} \sigma_{\mathsf{M}_s}^{M_s}(dx_s)\right\}
\sigma_{\mathsf{M}_{k-1}}^{M_{k-1}}(dx_{k-1})\times \\
&
\Bigg(\int_{\mathsf{M_k}}
\left[
g_k(x_k) \int_{\mathsf{M}_{k-1}}
f(x_{k-1},x_k)
\Bigg\{ \int_{\pmb{\mathsf{M}}_{k-2} }
\prod_{s=1}^{k-1} g_s(x_s)  f_s(x_{s-1},x_s)
\bigotimes_{s=1}^{k-2} \sigma_{\mathsf{M}_s}^{M_s}(dx_s)\right\}\times\\
&
\sigma_{\mathsf{M}_{k-1}}^{M_{k-1}}(dx_{k-1})
\Bigg]\sigma_{\mathsf{M}_k}^{M_k}(dx_k)\Bigg)^{-1}
\\
& =
\frac{
g_k(x_k)
\int_{\mathsf{M_{k-1}}}
f_k(x_{k-1},x_k)
\pi_{k-1}(x_{k-1})
\sigma_{\mathsf{M}_{k-1}}^{M_{k-1}}(dx_{k-1})
}
{
\int_{\mathsf{M_k}}
g_k(x_k)
\int_{\mathsf{M_{k-1}}}
f_k(x_{k-1},x_k)
\pi_{k-1}(x_{k-1})
\sigma_{\mathsf{M}_{k-1}}^{M_{k-1}}(dx_{k-1})
\sigma_{\mathsf{M}_k}^{M_k}(dx_k)
}
\end{align*}
where we have divided by
$$
\int_{\pmb{\mathsf{M}}_{k-1}}
\prod_{s=1}^{k-1} g_s(x_s)  f_s(x_{s-1},x_s)
\bigotimes_{s=1}^{k-1} \sigma_{\mathsf{M}_s}^{M_s}(dx_s)
$$
in the numerator and denominator to move from the second to the third line.
\end{proof}

\subsection{Proof of Proposition \ref{prop:prop2}}\label{app:app2}

\begin{proof}
We consider the case $k\geq 2$ as the case $k=1$ follows in a similar manner and we only give the proof when considering acceptance part of the kernel as the proof for the rejection part will follow in almost the same way.
In other words we are considering what happens to
$$
\int_{\mathbb{R}^{d_x}}\varphi(z')
\min\left\{1,
\frac{
p_{\mathtt{N}}(u_k^{\Delta}(\widetilde{z}',\epsilon))
\tfrac{1}{N}
\sum_{i=1}^N f_k(u_{k-1}^{\Delta}(\widetilde{z}_{k-1}^i,x), u_k^{\Delta}(\widetilde{z}',x))
}{
p_{\mathtt{N}}(u_k^{\Delta}(\widetilde{z},\epsilon))
\tfrac{1}{N}
\sum_{i=1}^N f_k(u_{k-1}^{\Delta}(\widetilde{z}_{k-1}^i,x), u_k^{\Delta}(\widetilde{z},x))
}
\right\} \widetilde{q}_k(\widetilde{z},\widetilde{z}')
\lambda_{d_x}(d\widetilde{z}').
$$
As $\Delta\downarrow 0$ we have that  for any $\widetilde{z}\in\mathbb{R}^{d_x}$ (recall
$z=(z,\overline{z})^{\top}$)
$$
u_k^{\Delta}(\widetilde{z}) \rightarrow
\begin{bmatrix}
z_k^{\star} \\
0
\end{bmatrix} +
\begin{bmatrix}
V_k^{\star} z  \\
\bar{z}
\end{bmatrix}
$$
this is because
$$
V_k^{\Delta} \rightarrow
\begin{bmatrix}
V_k^{\star} & 0  \\
0 & \mathtt{I}_{d_y}
\end{bmatrix}.
$$
Therefore we have that
$$
\min\left\{1,
\frac{
p_{\mathtt{N}}(u_k^{\Delta}(\widetilde{z}',\epsilon))
\tfrac{1}{N}
\sum_{i=1}^N f_k(u_{k-1}^{\Delta}(\widetilde{z}_{k-1}^i,x), u_k^{\Delta}(\widetilde{z}',x))
}{
p_{\mathtt{N}}(u_k^{\Delta}(\widetilde{z},\epsilon))
\tfrac{1}{N}
\sum_{i=1}^N f_k(u_{k-1}^{\Delta}(\widetilde{z}_{k-1}^i,x), u_k^{\Delta}(\widetilde{z},x))
}
\right\}
\rightarrow
$$
$$
\min\left\{1,
\frac{p_{\mathtt{N}}(\bar{z}')
\tfrac{1}{N}
\sum_{i=1}^N f_k(u_{k-1}^{\star}(z_{k-1}^i), u_k^{\star}(z'))
}{p_{\mathtt{N}}(\bar{z})
\tfrac{1}{N}
\sum_{i=1}^N f_k(u_{k-1}^{\star}(z_{k-1}^i), u_k^{\star}(z))
}
\right\}.
$$
The proof is then concluded via dominated convergence.
\end{proof}

\end{document}